\documentclass[a4paper,11pt]{article}
\usepackage{jcappub}
\usepackage[T1]{fontenc}
\usepackage[caption=false]{subfig}
\usepackage{hyperref}
\usepackage{amsmath}
\usepackage{mathtools}
\usepackage{tablefootnote}
\usepackage{soul}
\usepackage{ulem}
\usepackage{color}
\usepackage{float}
\usepackage{multirow}
\usepackage{comment}
\usepackage[usenames,dvipsnames,svgnames,table]{xcolor}

\def\mean#1{\left< #1 \right>}          % Mean value

\title{Testing gravity with gravitational waves $\times$ electromagnetic probes cross-correlations}

\author[a,b,c]{Giulio Scelfo,}
\emailAdd{giulio.scelfo@sissa.it}
\author[a,b,c]{Maria Berti,}
\emailAdd{mberti@sissa.it}
\author[e]{Alessandra Silvestri,}
\emailAdd{silvestri@lorentz.leidenuniv.nl}
\author[a,b,c,d]{Matteo Viel}
\emailAdd{viel@sissa.it}

\affiliation[a]{SISSA, Via Bonomea 265, 34136 Trieste, Italy}
\affiliation[b]{INFN, Sezione di Trieste, Via Bonomea 265, 34136 Trieste, Italy}
\affiliation[c]{IFPU, Institute for Fundamental Physics of the Universe, via Beirut 2, 34151, Trieste, Italy}
\affiliation[d]{INAF/OATS, Osservatorio Astronomico di Trieste, via Tiepolo 11, I-34143 Trieste, Italy}
\affiliation[e]{Institute Lorentz, Leiden University, PO Box 9506, Leiden 2300 RA, The Netherlands}

\abstract{In a General Relativistic framework, Gravitational Waves (GW) and Electromagnetic (EM) waves are expected to respond in the same way to the effects of matter perturbations between the emitter and the observer. A different behaviour might be a signature of alternative theories of gravity. In this work we study the cross-correlation of resolved GW events (from compact objects mergers detected by the Einstein Telescope, either assuming or excluding the detection of an EM counterpart) and EM signals (coming both from the Intensity Mapping of the neutral hydrogen distribution and resolved galaxies from the SKA Observatory), considering weak lensing, angular clustering and their cross term ($\mathrm{L \times C}$) as observable probes. Cross-correlations of these effects are expected to provide promising information on the behaviour of these two observables, hopefully shedding light on beyond GR signatures. We perform a Fisher matrix analysis with the aim of constraining the $\{\mu_0,\eta_0,\Sigma_0\}$ parameters, either opening or keeping fixed the background parameters $\{w_0,w_a\}$. We find that, although lensing-only forecasts provide significantly unconstrained results, the combination with angular clustering and the cross-correlation of all three considered tracers (GW, IM, resolved galaxies) leads to interesting and competitive constraints. This offers a novel and alternative path to both multi-tracing opportunities for Cosmology and the Modified Gravity sector.}

\begin{document}
\maketitle
	
\section{Introduction}\label{sec:intro}
Nowadays we can probe the Universe by means of different sorts of observables and through a large set of working or planned experiments. The newest window is given by Gravitational Waves (GWs), leading to the birth of the so-called Gravitational Waves Astronomy after the first detection of a Binary Black Hole merger by the LIGO/Virgo scientific collaboration~\cite{abbott:firstligodetection, abbott:firstligodetectionproperties} and anticipating a plethora of new detections~\cite{Abbott:O12,Abbott:O3}. All together with forthcoming experiments (such as the Einstein Telescope (ET)~\cite{Sathyaprakash:ET}, Cosmic Explorer~\cite{CE:2019}, LISA~\cite{amaro2017LISA}, KAGRA~\cite{somiya:kagra}, and LIGO-India~\cite{unnikhrishnan:indigo}), investigation of the Universe through this observation window is just at its promising beginning.

Another innovative technique is constituted by the Line Intensity Mapping (LIM, or simply IM), i.e., the measurement of the integrated emission from spectral lines from unresolved galaxies and
diffuse intergalactic medium (see e.g., references~\cite{kovetz17:lim,bernal2022line} for comprehensive reviews). IM surveys aim at scanning large portions of the sky in a relatively small amount of time by measuring the intensity
of a chosen emission line instead of resolving single galaxies. The result is a map of the underlying matter distribution, whose redshift information is finely accurate, thanks to the fact that the emission frequency of the line is known precisely. On the other hand, brightness temperature fluctuations reflect the distribution of underlying Large Scale Structure (LSS), as brighter signals are associated with denser regions. One of the most popular lines under study is the so-called 21 cm, emitted from the spin-flip transition of neutral hydrogen (HI), often studied in cross-correlation with galaxy surveys (see e.g.~\cite{Chang2010,Masui2013,Villaescusa+14:reio,Anderson2018,Wolz2021}). HI IM surveys are active or planned through experiments like MeerKAT~\cite{Santos+17,Wang+21:meerkat}, CHIME~\cite{bandura2014}, FAST~\cite{Hu2020}, BINGO~\cite{Battye2016}, Tianlai~\cite{Tianlai}, and HIRAX~\cite{Newburgh2016}. Particular interest is associated with the Square Kilometre Array Observatory (SKAO)~\cite{Braun2015:ska} due to the expected cosmological constraints it should bring~\cite{SKA_redbook,maartens:ska,Santos15:SKA}. 

Finally, we can find in (resolved) galaxy surveys another not novel but very powerful observation window with past, present and planned surveys/instruments shedding light on both Astrophysics and Cosmology (e.g., Euclid~\cite{laureijs2011euclid}, EMU~\cite{EMU}, DESI~\cite{desi}, SKAO~\cite{Braun2015:ska}, Vera Rubin Observatory (LSST)~\cite{LSST_DE}, JWST~\cite{JWST}, SPHEREx~\cite{spherex}, WFIRST~\cite{wfirst}, and several others).

All these experiments, targeting different observables, are producing a large amount of data, which will become more abundant with forthcoming experiments in the relatively near future. Given this variety, it is reasonable to explore the scientific opportunities that can arise from combining together different data-sets, i.e., studying the cross-correlation of different tracers of the underlying LSS. Indeed, in several published works cross-correlations between the LSS and the Cosmic Microwave Background (e.g.~\cite{nolta:2004, ho:correlation, hirata:correlation, raccanelli:crosscorrelation, raccanelli:radio, raccanelli:isw, Bianchini:2014dla, Bianchini:2015fiw, Bianchini:2015yly, Mukherjee:gwxcmb}), neutrinos (e.g.~\cite{fang:cross}), different LSS tracers (e.g.~\cite{Martinez:cross,Jain:cross,Yang:cross,Paech:cross}), IM (e.g.~\cite{Schmidt13:cross,Alonso15,Kovetz16:cross,Alonso16:cross,Wolz2016:cross,Pourtsidou16:cross,Pourtsidou16:cross_2,Raccanelli16:cross,Pourtsidou17:IM,Wolz17:cross,Alonso17:lssxim,Wolz18:cross,Cunnington18:lssxim}) or GWs (e.g.~\cite{Oguri:2016, raccanelli:pbhprogenitors, Scelfo18:gwxlss,Scelfo20:gws,namikawa:cross_ng,alonso:cross,Canas:sgwb,Calore:crosscorrelating,camera:gwlensing, Libanore+21, Scelfo:gwxim, Mukherjee:gwxlss1, Mukherjee:gwxlss2,Mukherjee:sgwb,canas2021gaus,mukherjee22:H0,Cigarran22}) have been studied.

In this work, we explore the cross-correlation between GW events from resolved Compact Objects (CO) mergers and electromagnetic (EM) signals coming from luminous tracers, such as the IM of the 21 cm line and resolved galaxies. We consider the ET instrument for the first observable, and SKAO for the latter ones. We exploit these different probes with the aim of testing the possibility of gravity theories alternative to General Relativity (GR). Indeed, once a GW or an EM signal is emitted from a source, cosmic structures between the origin and the observer interfere through distortion effects under the form of magnifications (or de-magnifications). In a standard GR framework, these effects are expected to act in the same way on GW and EM waves, whereas different imprints may be a signal of deviations from GR, indicating the need for Modified Gravity (MG) theories. Consequently, cross-correlations between these distortion effects on these probes should highlight potential MG behaviours and help set constraints on related physical parameters. Thus, our main observable is the lensing power spectrum (both in auto and cross-tracers correlations). Subsequently, we also combine it with data from angular clustering power spectra, in order to test the improvement brought by the merger of different observational probes. This avenue of cross-correlating GW and EM signals to test gravity was already explored in the literature (see e.g., references~\cite{Mukherjee:gwxlss1,Mukherjee:gwxlss2,Baker:2021,Mukherjee21:dark}). We expand on previous works by considering a larger variety of tracers (GWs, resolved galaxies and IM, eventually simultaneously) and different probes combinations (lensing, angular clustering and their cross-correlation).

This manuscript is structured as follows: in section \ref{sec:formalism} we describe our methodology, presenting the treated probes (weak lensing, angular clustering, and their cross-term) in section \ref{sec:formalism_Cls} and the adopted Fisher analysis formalism in section \ref{sec:formalism_Fisher}; in section \ref{sec:tracers} we introduce and characterize the considered tracers (GWs, IM and resolved galaxies); in section \ref{sec:MG} we introduce the tested MG parametrization; in section \ref{sec:forecasts} we present our forecasts on the relevant MG parameters and in section \ref{sec:conclusions} we draw our conclusions.

\section{Methodology}\label{sec:formalism}
In this section we describe the observables considered and the adopted methodology. In section \ref{sec:formalism_Cls} we characterize our observables: the angular power spectra for weak lensing and angular clustering (and their cross term). In section \ref{sec:formalism_Fisher} we describe the Fisher formalism on which we rely.
\subsection{Observables: angular power spectra (in $\Lambda$CDM)}
\label{sec:formalism_Cls}
The observables we consider are the angular power spectra $C_\ell$s for two different probes: weak lensing (denoted as L) and angular clustering (denoted as C), with the addition of the cross-term ($\mathrm{L \times C}$). Given two tracers \{X,Y\} (e.g., GW events, galaxies, IM) associated to two different redshift bins $\{z_i,z_j\}$, we define the power spectra of their cross-correlation as $C^{\rm X_i,Y_j}_{\Gamma \Theta}(\ell)$, with $\Gamma,\Theta$ indicating the considered probe (e.g., L or C). We make use of the flat-sky and Limber approximations, which are accurate at 10\% for $\ell = 4$, 1\% for $\ell = 14$, and less than 0.1\% for $\ell > 45$~\cite{Shear_full_sky-Kilbinger+17}. In the following, we characterize the power spectra for the considered probes.

\begin{itemize}
    \item \textbf{Weak lensing (L)}. The characterization and physical meaning of this observable depends on the tracer that we take into account. For what concerns resolved galaxies, it describes the physical effect of distortion of their shape due to the inhomogeneous distribution of matter between
the objects and the observer. It is often referred to as \textit{cosmic shear} (see e.g.,~\cite{Bartelmann:WL,Hoekstra:WL}). It is given by the sum of three different terms: the proper cosmological signal ($\gamma \gamma$ term) and the two intrinsic alignment terms ($\gamma$I and II terms). The latter ones consider that observed galaxies are usually already characterized by an intrinsic ellipticity, which should be taken into account when estimating the shear due to weak lensing only. The three terms can be written as (see e.g.,~\cite{joachimi2015:IA}):
    \begin{eqnarray}
    C_{\gamma\gamma}^{\rm X_iY_j}(\ell) &=& \int_0^\infty \frac{d z \ c}{H(z)} \  \frac{W_\gamma^{\rm X_i}(z) \ W_\gamma^{\rm Y_j}(z)}{\chi^2(z)} \ P_\mathrm{mm}\left(\frac{\ell}{\chi(z)},z\right)
    \label{eq:weak_lensing_gg}
    \\
    C_{\gamma \mathrm I}^{\rm X_iY_j}(\ell) &=& \int_0^\infty \frac{d z \ c}{H(z)} \  \frac{W_\gamma^{\rm X_i}(z) \ W_\mathrm{IA}^{\rm Y_j}(z) + W_\mathrm{IA}^{\rm X_i}(z) \ W_\gamma^{\rm Y_j}(z)}{\chi^2(z)} \times \nonumber \\
    &&
    \times \ \mathcal F_\mathrm{IA}(z) \ P_\mathrm{mm}\left(\frac{\ell}{\chi(z)},z\right)
    \label{eq:weak_lensing_gI}
    \\
    C_\mathrm{II}^{\rm X_iY_j}(\ell) &=& \int_0^\infty \frac{d z \ c}{H(z)} \  \frac{W_\mathrm{IA}^{\rm X_i}(z) \ W_\mathrm{IA}^{\rm Y_j}(z)}{\chi^2(z)} \ \mathcal F_\mathrm{IA}^2(z) \  P_\mathrm{mm}\left(\frac{\ell}{\chi(z)},z\right),
    \label{eq:weak_lensing_II}
    \end{eqnarray}
    
    where $c$ is the speed of light, $H(z)$ is the Hubble parameter, $\chi(z)$ is the comoving distance, $P_{\rm mm}$ is the matter power spectrum and the window functions are given by:
    \begin{eqnarray}
    W_\gamma^{\rm X_i}(z) &=& \frac{3}{2} \Omega_\mathrm m \ \frac{H_0^2}{c^2} \chi(z) (1+z) \int_z^\infty d x \ n_{\rm X_i}(x) \ \frac{\chi(x)-\chi(z)}{\chi(x)}
    \label{eq:weak_lensing_window_g}
    \\
    W_\mathrm{IA}^{\rm X_i}(z) &=& n_{\rm X_i}(z) \ \frac{H(z)}{c},
    \label{eq:weak_lensing_window_IA}
    \end{eqnarray}
    
    where $n_{\rm X_i}$ is the redshift distribution of the considered tracer and the intrinsic alignment kernel $\mathcal F_\mathrm{IA}$ is modeled through the extended non-linear alignment model:
    \begin{equation}
    \mathcal F_\mathrm{IA}(z) = -\frac{A_\mathrm{IA} \mathcal C_1 \Omega_\mathrm m}{D_1(z)} (1+z)^{\eta_\mathrm{IA}} \left(\frac{\mean{L}(z)}{L_*(z)}\right)^{\beta_\mathrm{IA}},
    \end{equation}
    with $\mathcal C_1 = 0.0134$, $D_1(z)$ is the linear growth factor and the intrinsic alignment parameters have fiducial values $\{A_\mathrm{IA}, \eta_\mathrm{IA}, \beta_\mathrm{IA}\} = \{1.72,-0.41, 2.17\}$.
    Finally, $\frac{\mean{L}(z)}{L_*(z)}$ is the mean luminosity of the sample in units of the typical luminosity at a given redshift. Here, we use the same specification used for Euclid~\cite{blanchard2020euclid}, both for ease of comparison with similar studies and also under the assumption that the galaxies observed by SKAO will display a similar redshift evolution of their luminosity. However, we note that this assumption must be explicitly checked, by performing an analysis on actual observations, as in reference~\cite{joachimi11}.
    
   Equations \ref{eq:weak_lensing_gg} - \ref{eq:weak_lensing_II} can be summed up to give the lensing power spectrum
    \begin{equation}
    C_\mathrm{LL}^{\rm X_iY_j}(\ell) =
    \int_0^\infty \frac{d z \ c}{H(z)} \  \frac{W_\mathrm L^{\rm X_i}(z) \ W_\mathrm L^{\rm Y_j}(z)}{\chi^2(z)} \ P_\mathrm{mm}\left(\frac{\ell}{\chi(z)},z\right),
    \label{eq:weak_lensing_LL}
    \end{equation}

    where 
    \begin{equation}
    W_\mathrm L^{\rm X_i}(z) = W_\gamma^{\rm X_i}(z) + \mathcal F_\mathrm{IA}(z) \ W_\mathrm{IA}^{\rm X_i}(k,z).
    \label{eq:weak_lensing_window_L}
    \end{equation}
    In the case of GW events we do not have an intrinsic shape that undergoes cosmic shear, so the intrinsic alignment term is not present. Indeed, in this case, the propagation of the gravitational wave in the presence of a matter distribution leads to magnification in the strain signal $h\left(f\right)$:
    \begin{equation}
    h\left(f\right)=\mathcal{Q}( { \alpha}) \sqrt{\frac{5}{24}} \frac{G^{5 / 6} \mathcal{M}^2\left(f \mathcal{M}\right)^{-7 / 6}}{c^{3 / 2} \pi^{2 / 3} d_L} e^{i \phi},
    \end{equation}
    where $f$ is the frequency, $\mathcal{Q}({ \alpha})$ is a function of the angles describing the position and orientation of the binary, $\mathcal{M}$ is the chirp mass of the binary system, $d_L$ is the luminosity distance of the source and $G$ is the gravitational constant.
    What one can measure is an alteration in the measured GW strain
    $\tilde{h}\left(\hat{r},f\right)=h\left(f\right)\left[1+\kappa(\hat{r})\right]$, where $\hat{r}$ describes the position of the source and $\kappa(\hat{r})$ is the lensing convergence, related to the angular power spectra as  $C_\mathrm{LL}(\ell) = \langle \kappa_{\ell m} \kappa_{\ell' m'}\rangle \delta_{\ell \ell'} \delta_{m m'} $. We refer the interested reader to e.g., references~\cite{Takahashi2006,Laguna2010,Cutler2009,camera:gwlensing,Congedo19:WL,Bertacca:GWDL,Mukherjee:gwxcmb,Mukherjee:gwxlss1,mpetha22:WL} for further details.
    
    Finally, although IM (by definition) is a probe that does not provide resolved galaxies, we can still describe the effects of weak lensing as a magnification received by the observer (see e.g., references~\cite{Pourtsidou:IMlensing1, Pourtsidou:IMlensing2} for additional details). As one would expect, also in this case the IA term is not present ($\mathcal{F}_\mathrm{IA}^{\rm IM}(z)=0$).

    \item \textbf{Angular clustering (C)}. Our tracers can also be used to estimate the clustering as a function of the separation angle (or equivalently the multipoles):
    \begin{equation}
    C_\mathrm{CC}^{\rm X_iY_j}(\ell) = \int_0^\infty \frac{d z \ c}{H(z)} \  \frac{W_\mathrm C^{\rm X_i}\left(\frac{\ell}{\chi(z)},z\right) \ W_\mathrm C^{\rm Y_j}\left(\frac{\ell}{\chi(z)},z\right)}{\chi^2(z)} \ P_\mathrm{mm}\left(\frac{\ell}{\chi(z)},z\right),
    \label{eq:weak_lensing_GG}
    \end{equation}

    where the window function for clustering is given by
    \begin{equation}
    W_\mathrm C^{\rm X_i}(k,z) = b_{\rm X}(k,z) \ n_{\rm X_i}(z) \frac{H(z)}{c}
    \label{eq:weak_lensing_window_G}
    \end{equation}
    and $b_{\rm X}(k,z)$ is the bias parameter for tracer $X$, describing the relation between the tracer and the underlying matter distribution (see e.g.,~\cite{Kaiser:bias, Bardeen:bias, Mo:smallhalosbias, matarrese:clusteringevolution, dekel:stochasticbiasing, benson:galaxybias, peacock:halooccupation, Desjacques:bias}).
    
    We apply this formalism to all tracers considered in this work.

    \item \textbf{Lensing $\times$ Clustering ($\mathrm{L \times C}$)}. Finally, the cross-correlation $\mathrm{L \times C}$ between weak lensing and angular clustering of two tracers can be expressed as
    \begin{equation}
    C_\mathrm{CL}^{\rm X_iY_j}(\ell) = \int_0^\infty \frac{d z \ c}{H(z)} \ \frac{W_\mathrm C^{\rm X_i}\left(\frac{\ell}{\chi(z)},z\right) \ W_\mathrm L^{\rm Y_j}(z)}{\chi^2(z)} \ P_\mathrm{mm}\left(\frac{\ell}{\chi(z)},z\right).
    \label{eq:weak_lensing_GL}
    \end{equation}
    Essentially, it is given by the combination of a Lensing window function with a Clustering one.
    \end{itemize}

\subsection{Fisher analysis}
\label{sec:formalism_Fisher}
In this work, we make use of the Fisher matrix analysis, which we briefly sketch in this section.
Assuming again two tracers \{X,Y\} (e.g., GW events, galaxies, IM), we divide the total redshift interval surveyed in $N_{\mathrm{bins}}^{\mathrm{X}}$ bins, with amplitude $\Delta z^{\mathrm{X}}$ for tracer X, and in $N_{\mathrm{bins}}^{\mathrm{Y}}$ redshift bins with amplitude $\Delta z^{\mathrm{Y}}$ for tracer Y.

\begin{figure}
	\centering
	\includegraphics[width=0.7\linewidth]{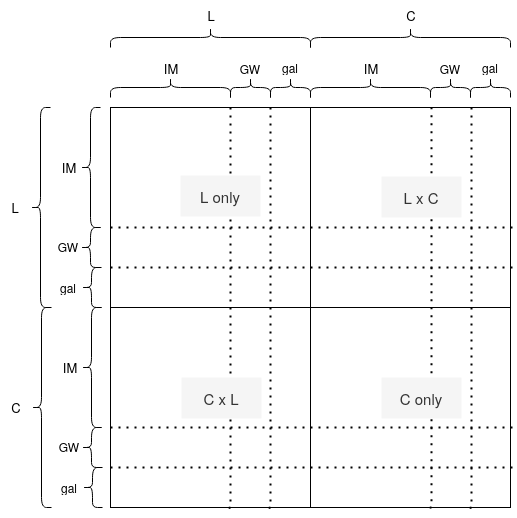}
	\caption{Sketch for the $\mathcal{C}_\ell$ matrix in the case of all probes (L only, C only, and $\mathrm{L \times C}$) and three tracers (GW, IM, gal) considered simultaneously.}
	\label{fig:Cl_matrix_total}
\end{figure}

Considering the observed power spectra $\tilde{C}_\ell$s for a specific probe (L only, C only or $\mathrm{L \times C}$, which we do not explicitate throughout this section) and a generic set of parameters $\{\theta_n\}$ for the Fisher analysis, we can organize our data in the (symmetric) matrix $\mathcal{C}_\ell$ as
\begin{equation}
\mathcal{C}_\ell=
\begin{bmatrix}
\tilde{C_\ell}^{\mathrm{X\,X}}(z_1^{\mathrm{X}},z_1^{\mathrm{X}}) & ... & \tilde{C_\ell}^{\mathrm{X\,X}}(z_1^{\mathrm{X}},z_N^{\mathrm{X}}) & \tilde{C_\ell}^{\mathrm{X\,Y}}(z_1^{\mathrm{X}},z_1^{\mathrm{Y}}) & ... &
\tilde{C_\ell}^{\mathrm{X\,Y}}(z_1^{\mathrm{X}},z_N^{\mathrm{Y}}) \\
& ... & \tilde{C_\ell}^{\mathrm{X\,X}}(z_2^{\mathrm{X}},z_N^{\mathrm{X}}) & \tilde{C_\ell}^{\mathrm{X\,Y}}(z_2^{\mathrm{X}},z_1^{\mathrm{Y}}) & ... &
\tilde{C_\ell}^{\mathrm{X\,Y}}(z_2^{\mathrm{X}},z_N^{\mathrm{Y}})\\
& ... & $\vdots$ & $\vdots$ & ... & $\vdots$\\
& & \tilde{C_\ell}^{\mathrm{X\,X}}(z_N^{\mathrm{X}},z_N^{\mathrm{X}}) & \tilde{C_\ell}^{\mathrm{X\,Y}}(z_N^{\mathrm{X}},z_1^{\mathrm{Y}})  & ... &
\tilde{C_\ell}^{\mathrm{X\,Y}}(z_N^{\mathrm{X}},z_N^{\mathrm{Y}}) \\
&  &  & \tilde{C_\ell}^{\mathrm{Y\,Y}}(z_1^{\mathrm{Y}},z_1^{\mathrm{Y}}) & ... &
\tilde{C_\ell}^{\mathrm{Y\,Y}}(z_1^{\mathrm{Y}},z_N^{\mathrm{Y}})\\
& & & & ... & $\vdots$ \\
&  & & & & \tilde{C_\ell}^{\mathrm{Y\,Y}}(z_N^{\mathrm{Y}},z_N^{\mathrm{Y}})\\
\end{bmatrix},
\label{eq:C_matrix_multi}
\end{equation}  
The matrix $\mathcal{C}_\ell$ has dimensions of $(N_{\mathrm{bins}}^{\mathrm{X}}+N_{\mathrm{bins}}^{\mathrm{Y}}) \times (N_{\mathrm{bins}}^{\mathrm{X}}+N_{\mathrm{bins}}^{\mathrm{Y}})$. Note that in general $z_i^{\mathrm{X}} \neq z_i^{\mathrm{Y}}$, since the two tracers may be distributed among different bins. We stress again that the tilde symbol stands for \textit{observed} $C_\ell$s. It is trivial to expand the above matrix to the case in which a third tracer Z is considered at the same time. In this case, the matrix would be accordingly expanded with all XZ, YZ and ZZ correlations and would have dimensions of $(N_{\mathrm{bins}}^{\mathrm{X}}+N_{\mathrm{bins}}^{\mathrm{Y}}+N_{\mathrm{bins}}^{\mathrm{Z}}) \times (N_{\mathrm{bins}}^{\mathrm{X}}+N_{\mathrm{bins}}^{\mathrm{Y}}+N_{\mathrm{bins}}^{\mathrm{Z}})$. The three tracers case is also explored in this work (see sections \ref{sec:tracers} and \ref{sec:forecasts}).
Equation \eqref{eq:C_matrix_multi} refers to the case in which just one probe is taken into account (L only, C only, or $\mathrm{L \times C}$). When all three probes are considered simultaneously for a forecast, the global $\mathcal{C}_\ell$ matrix will be made of 4 different sub-matrices like the one in equation \eqref{eq:C_matrix_multi}: one for L only, one for C only, and two for $\mathrm{L \times C}$. We provide in figure \ref{fig:Cl_matrix_total} a sketch of the global $\mathcal{C}_\ell$ matrix in the case of all probes and three tracers (GW, IM, gal as described in section \ref{sec:tracers}). Its dimensions are $2 (N_{\mathrm{bins}}^{\mathrm{IM}}+N_{\mathrm{bins}}^{\mathrm{GW}}+N_{\mathrm{bins}}^{\mathrm{gal}}) \times  2 (N_{\mathrm{bins}}^{\mathrm{IM}}+N_{\mathrm{bins}}^{\mathrm{GW}}+N_{\mathrm{bins}}^{\mathrm{gal}})$.

The $\mathcal{C}_\ell$ matrix is then used to compute the Fisher matrix elements as
\begin{equation}\label{eq:Fisher_matrix}
F_{\alpha \beta}=f_\mathrm{sky}\sum_\ell \frac{2\ell+1}{2} \mathrm{Tr}\left[\mathcal{C}_\ell^{-1} (\partial_\alpha \mathcal{C}_\ell)\mathcal{C}_\ell^{-1}(\partial_\beta \mathcal{C}_\ell)\right],
\end{equation}
where $\partial_{\alpha}$ indicates the partial derivative with respect to the parameter $\theta_{\alpha}$ and $f_\mathrm{sky}$ is the fraction of the sky covered by the intersection of the considered surveys. The Fisher-estimated marginal error on the parameter $\theta_{\alpha}$ is given by $\sqrt{(F^{-1})_{\alpha \alpha}}$.  According to the Cramér-Rao bound, the quantity $\sqrt{(F^{-1})_{\alpha \alpha}}$ provides the smallest expectable error for a ``real-life'' experiment, setting a lower bound to its estimate (and having the equality only in the case of gaussian likelihood and errors). Fisher approach may not always be the most accurate method to adopt since instrumental/observational systematic errors and/or the parameter posterior may not be gaussianly distributed. Still, it remains a simple and fast method to yield forecasts for designed experiments, providing reasonable results, especially for a first estimate. The novelty of this work allows us to adopt a Fisher formalism while considering its estimates informative enough to bring meaningful and reliable conclusions, although different techniques (such as Markov-Chain Monte-Carlo~\citep{gilks:1995}) may be suggested for further investigation. We refer the interested reader to references~\cite{Bellomo20:multiclass, Bernal20:multiclass} for further discussion about Fisher analysis and the impact of several approximations therein and in the observables considered.

\section{Tracers}
\label{sec:tracers}
In this section, we characterize the considered tracers. In table \ref{tab:tracers} we summarize their redshift dependent specifics (binning, redshift range, etc.).

\begin{table}[]
\centering
\begin{tabular}{|c|c|c|c|}
\hline
\textbf{\:\:\:\: Tracer \:\:\:\:}    & {$\rm{GW}^{\rm bright}$ (ET)}     & \:\: {$\rm{GW}^{\rm dark}$ (ET) $\&$ gal (SKAO)} \:\:           & \:\: {IM (SKAO)} \:\:                \\ \hline
\textbf{z range}    & [0.5-2.5]    & \multicolumn{2}{c|}{{\:\:\:\:\:\:\:\:\:\:\:\:\:\:\:\:\:\:\:\:\:\:\:\:\:\:\:\:\:\:[}0.5-3.5{]} } \\ \hline
$\mathbf{N_{\rm \bf bins}}$ & 8 & 3             & 30                 \\ \hline
$\mathbf{\Delta z}$    & 0.25  & 1.0           & 0.1                \\ \hline
\end{tabular}
\caption{Specifics for the considered tracers: redshift range, number of redshift bins $N_{\rm bins}$ and bin width $\Delta z$.}
\label{tab:tracers}
\end{table}

\subsection{Gravitational Waves}\label{sec:tracer_GW}
\begin{figure}
	\centering
	\includegraphics[width=1.0\linewidth]{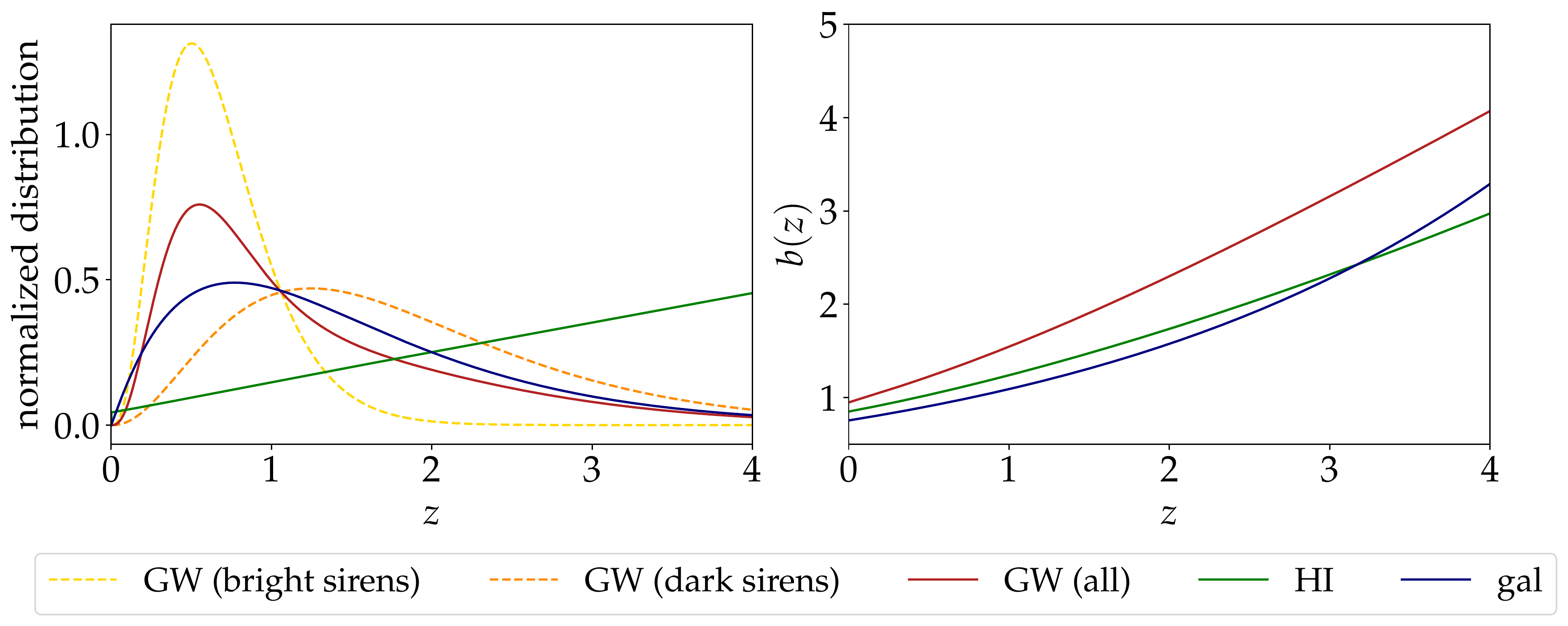}
	\caption{Specifics for the considered tracers: HI (green), resolved SKAO galaxies (blue), detected GW events from ET for dark sirens (orange), bright sirens (yellow), and both combined (dark red). \textit{Left panel}: normalized redshift distributions (number counts for GWs and resolved galaxies, mean brightness temperature $T_b(z)$ for HI). \textit{Right panel}: biases.}
	\label{fig:tracers_specs}
\end{figure}

We consider GW events from compact objects resolved mergers (BHBH, BHNS, and NSNS) detected by the Einstein Telescope (ET) experiment, as planned in~\cite{Sathyaprakash:ET}. We treat two categories of GW events, depending on whether they can be associated with an EM counterpart:
\begin{itemize}
    \item {\bf Dark sirens}: they are not accompanied by an EM follow-up. We treat BHBH and BHNS mergers as dark sirens and consider $N_{\mathrm{bins}}^{\mathrm{GW^{dark}}}=3$ redshift bins with width $\Delta z^{\mathrm{{GW}^{dark}}}=1.0$ in the redshift range $[0.5-3.5]$. We choose large redshift bins to take into account the poor redshift localization of this kind of sources. Given the lack of an EM counterpart, their angular resolution is limited by the capabilities of the considered GW instrument, which we set to $\ell_{\rm max}=100$~\cite{Sathyaprakash:ET}.
    \item {\bf Bright sirens}: the GW emission is associated with an EM counterpart. This helps not only in improving the angular localization of the emitting source but provides also extra information in the MG context, due to the fact that GWs and EM waves might behave differently depending on the MG model under consideration (see section \ref{sec:MG} for further details). We treat NSNS mergers as bright sirens and consider $N_{\mathrm{bins}}^{\mathrm{GW^{bright}}}=8$ redshift bins with width $\Delta z^{\mathrm{{GW}^{bright}}}=0.25$ in the redshift range $[0.5-2.5]$. This is motivated by the $z$-uncertainty behaviour for NSNS binaries $\delta z / z \approx 0.1 \: z$~\cite{Safarzadeh:NSNSerr}, making our choice quite conservative at lower redshifts. Since the detection of an EM follow-up can help in significantly improving the angular localization of the sources, it allows us to push our analysis to a higher $\ell_{\rm max}$. We set $\ell_{\rm max}=300$ for bright sirens, which appears to be a conservative estimate for this type of experiments (see e.g.~\cite{Mukherjee:gwxcmb,Mukherjee:gwxlss1,Balaudo22}), furthermore allowing us to avoid non-linearities in the power spectra modeling. We comment on the impact of the choice of $\ell_{\rm max}$ in section \ref{sec:forecasts}.
\end{itemize}
Prescriptions to describe the redshift evolution of the GW tracers and their bias parameter are taken from references~\cite{Boco19:gws, Scelfo20:gws} and provided in figure \ref{fig:tracers_specs}. These specifics predict a detection of $\sim 2.2 \cdot 10^4$ BHBH+BHNS mergers and $\sim 1.4 \cdot 10^4$ NSNS mergers in the corresponding redshift intervals (for $T^{\rm GW}_{\rm obs} = 1$yr and $f_{\rm sky}=0.5$). The GW events bias is evaluated through an abundance matching technique (see e.g.,~\cite{aversa+15}), linking the luminosity/SFR of each host galaxy to the mass of the hosting dark matter halo, eventually matching the bias of the associated halo to a galaxy with given SFR. Lastly, characterizing COs mergers with the same bias of their host galaxies, the final bias expression is estimated by taking into account which galaxy types give the biggest contribution to the observed merger rate proportionally. For further details on the GW bias estimate procedure we refer the interested reader to~\cite{Boco19:gws, Scelfo20:gws} and references therein.

Given a theoretical predicted value for the $C_{\ell}$s under study (computed with \texttt{COLIBRI}\footnote{See \url{https://github.com/GabrieleParimbelli/COLIBRI}.}, which we modified to extend it to the multi-tracing case), the observed power spectra $\tilde{C}_{\ell}$s are characterized by the presence of extra noise terms: $\tilde{C}_{\ell}^{\rm GW,X}(z_i,z_j) = C_{\ell}^{\rm GW,X}(z_i,z_j) + C_{\ell}^{\rm N,GW}(z_i)$. In the case of GWs-related power spectra, following e.g.~\cite{Mukherjee:gwxlss1}, we assume that:
\begin{equation}
    C_{\ell}^{\rm N,GW}(z_i) = \frac{1}{n_{\rm GW}} e_{d_L}^2 \exp \frac{\ell^2 \theta_{\rm min}^2}{8 \ln2},
\end{equation}
where $n_{\rm GW}$ is the number density of sources in the considered redshift bin $z_i$, $\theta_{\rm min}$ is the sky localization area of the gravitational wave sources, and $e_{d_L} \sim 3/{\rm SNR}$ is the relative error on the luminosity distance estimation (see e.g.~\cite{DelPozzo:GW,Calore:crosscorrelating}), where the average value of the Signal-to-Noise ratio (SNR) estimate for detected GWs events is derived by results from reference~\cite{Scelfo20:gws} and takes the values of SNR=8.4 (15.4) for bright (dark) sirens. We assume that this shot-noise/beam noise term affects all the probes considered in this work (i.e., L, C, L$\times$C).

\subsection{Neutral Hydrogen Intensity Mapping}\label{sec:tracer_HI}

We consider the forecasted HI distribution given by the SKA-Mid intensity mapping survey~\cite{SKA_redbook, Dewdney:ska, maartens:ska}. We consider the redshift range $[0.5-3.5]$, divided in bins of width $\Delta z^{\mathrm{IM}}=0.1$, for a total of $N_{\mathrm{bins}}^{\mathrm{IM}}=30$ redshift bins. This is expected to be around the optimal redshift range for the SKA-Mid survey~\cite{SKA_redbook}. The HI mean brightness temperature redshift evolution and the bias are taken from references~\cite{Battye:T_HI,Spinelli20:HI} and provided in figure \ref{fig:tracers_specs}. The HI bias prescription derives from the outputs of a semi-analytical model for galaxy formation explicitly incorporating a treatment of neutral hydrogen and are in agreement with results of \cite{Villaescusa+18:ingr} based on Illustris TNG hydro-dynamical simulations.

Noise sources for IM are the result of contributions from different elements, described as follows:
\begin{itemize}
    \item {\bf Beam effects}: the relation between theoretical $C_\ell^{XY}$ and the observed $\tilde{C}_\ell^{XY}$ is:
\begin{equation}\label{eq:IMxIM_obs}
\tilde{C}_\ell^{\rm IM,IM}(z_i,z_j) = \mathcal{B}(z_i)\mathcal{B}(z_j)C_\ell^{\rm IM,IM}(z_i,z_j)+C_{\ell}^{\mathrm{N,IM}}
\end{equation}
and
\begin{equation}
\tilde{C}_\ell^{\rm IM,X}(z_i,z_j) = \mathcal{B}(z_i)C_\ell^{\rm IM,X}(z_i,z_j)
\end{equation}
where the $\mathcal{B}^X(z_i)$ describes the suppression of the signal at scales smaller than the FWHM of the beam  $\theta_{B}$. In single-dish configuration $\theta_{B} \sim  1.22 \lambda/D_{d}$, implying a stronger suppression of the signal at lower frequencies:
\begin{equation}
    \mathcal{B}(z_i)=
    \exp[-\ell(\ell+1)(\theta_B(z_i)/\sqrt{16\ln2})^2].
\end{equation}
The beam term affects all probes considered (L, C, L$\times$C).

\item{\bf Foreground noise}: 
IM data analysis has to deal with the delicate cleaning procedure of the signal from the bright foreground emission (see e.g., references~\cite{Switzer+2013,Alonso14:fogs,Carucci:2020enz,Cunnington:2020njn,Matshawule2020,Wolz2021,Soares:2021}). Although modeling the foregrounds is beyond the scope of this work, we need to take into account the residual error that could be expected after a foreground removal procedure. Following reference~\cite{Scelfo:gwxim}, we model the foreground-cleaning related noise term as
\begin{equation}\label{eq:noise_fog}
C_\ell^{\mathrm{fg}} = K^{\mathrm{fg}}\cdot F(\ell),
\end{equation}
where $K^{\mathrm{fg}}$ is an overall normalization constant determining the overall amplitude of the residual foregrounds related errors and is given by an average value of all the $C_\ell^{\mathrm{IM,IM}}(z_i,z_j)$ components:
\begin{equation}
K^{\mathrm{fg}} = \left\langle C_\ell^{\mathrm{IM,IM}}(z_i,z_j)\right\rangle.
\end{equation} 
The function $F(\ell)$ encodes the scale-dependence, described by
\begin{equation}\label{eq:F_fid}
F(\ell) = \dfrac{1}{f_{\mathrm{sky}}} A e^{b \ell^c},
\end{equation}
with a stronger error at larger scales. With the chosen numerical values ($A\sim0.129, \: b\sim-0.081, \: c\sim0.581$) the error is around 12\% at $\ell\sim 2$ and 4\% at $\ell \sim 100$ (for $f_{\mathrm{sky}}=1.0$). This term affects all probes (L, C, L$\times$C), but only $\mathrm{IM \times IM}$ terms (for all redshift bins combinations).
\item{\bf Instrumental noise}: the noise angular power spectrum for the experiment setup under study (single dish mode~\cite{SKA_redbook, Santos+17} with an ensemble of $N_d$ dishes) writes as (see e.g.,~\cite{Bull15:IM,Santos15:SKA,Santos+17}):
\begin{equation}\label{eq:noise_IM}
C_{\ell}^{\mathrm{instr}}(z_i)= \sigma_T^2 \theta_B^2 \approx \Biggl(\frac{T_{\mathrm{sys}}}{T_b(z_i)\sqrt{n_{\mathrm{pol}} B t_{\mathrm{obs}} N_d}} \sqrt{\dfrac{S_{\mathrm{area}}}{\theta_{\mathrm{B}}^{2}}}\dfrac{1}{T_b(z_i)}\Biggl)^2\theta_{B}^2, 
\end{equation}
where the single-dish rms noise temperature $\sigma_{T}$ is given by
\begin{equation}
\sigma_{T} \approx \frac{T_{\mathrm{sys}}}{\sqrt{n_{\mathrm{pol}} B t_{\mathrm{obs}}}} \frac{\lambda^{2}}{\theta_{\mathrm{B}}^{2} A_{e}} \sqrt{S_{\mathrm{area}} / \theta_{\mathrm{B}}^{2}} \sqrt{\frac{1}{N_{\mathrm{d}}}}
\end{equation}
and the other parameters involved are (according to SKA-Mid prescriptions): $T_{\rm sys} = 28K$ for the system temperature, $B=20\cdot 10^6 Hz$ for the bandwidth, $t_0 = 5000 h = 1.8 \cdot 10^{7} s$ for the observation time, $N_d=254$ for the total number of dishes, $S_{\mathrm{area}}=20000 deg^2$ for the total surveyed area, $A_e=140 m^2$, $D_d=15m$ and $S_{\mathrm{area}}=20000 \: deg^2 \sim f_{\rm sky}=0.5$ for the reference sky coverage~\cite{Santos15:SKA}. $T_b(z_i)$ is the mean brightness temperature at the center of the redshift bin and it acts as a normalization factor to retrieve a dimensionless $C_{\ell}$. This noise component affects all the probes considered in this work (L, C, L$\times$C) but it is de-correlated among different bins, affecting only IM auto-correlations.
\item{\bf Lensing reconstruction error}: references such as~\cite{Pourtsidou:IMlensing1,Pourtsidou:IMlensing2} model an extra scale independent noise contribution, due to inaccuracies in the reconstruction of the signal. Since it should affect scales smaller than our $\ell= \mathcal{O}(100)$ cut-off, we opt for not taking it into account. For sake of completeness, we checked that artificially introducing a noise term overcoming the observed signal at around 2/3 of the explored angular range, would worsen our forecasts by $\sim 15-20 \%$ or less. Still, let us stress again that the actual scales at which this noise is supposed to dominate start from around $\ell \sim \mathcal{O}(100)$, safely allowing us to neglect this term.
\end{itemize}

\subsection{Galaxies}\label{sec:tracer_gal}
We consider SKAO radio-galaxies distributed following the T-RECS catalog~\cite{Bonaldi:TRECS} for SKAO (radio continuum survey with $5 \mu JY$ detection threshold for $z<5$). We consider $N_{\mathrm{bins}}^{\mathrm{g}}=3$ redshift bins with width $\Delta z^{\mathrm{g}}=1.0$ in the redshift range $[0.5-3.5]$. Their redshift distribution and bias are provided in figure \ref{fig:tracers_specs} (see e.g., reference~\cite{Scelfo18:gwxlss} for further details). The galaxy bias formulation relies on outputs from the $S^3$ simulation \cite{wilman2008}. We model noise sources for SKAO radio galaxies as follows:
\begin{itemize}
    \item {\bf Shot noise}: the shot noise term affects only the Clustering probe and reads as
    \begin{equation}
        C_{\ell}^{\rm N,g} = C_{\ell}^{\rm shot,g} = \dfrac{1}{n_g},
    \end{equation}
    where $n_g$ is the source number density in the considered redshift bin. This term affects only ${\rm g(z_i) \times g(z_i)}$ terms (same tracer and same $z$ bin).
    \item {\bf Shape noise}: this term affects only the Lensing probe and it encodes the intrinsic ellipticity of observed galaxies, which may bias results if not taken into account. It reads as
    \begin{equation}
        C_{\ell}^{\rm N,g} = C_{\ell}^{\rm shape,g} = \dfrac{\gamma^2}{n_g}
    \end{equation}
    where $\gamma=0.3$ is the intrinsic shear term~\cite{Sprenger2019:sigmaz}. This term affects only ${\rm g(z_i) \times g(z_i)}$ terms (same tracer and same $z$ bin).
    \item {\bf Shot $\times$ shape noise}: being the $L \times C$ probe term made of the contribution of both Lensing and Clustering, we model its noise contribution as a mixture of the shot and shape noises affecting Clustering and Lensing respectively. It reads as
    \begin{equation}
        C_{\ell}^{\rm N,g} = \sqrt{\bigl(C_{\ell}^{\rm shot,g}\bigl)^2 + \bigl(C_{\ell}^{\rm shape,g}\bigl)^2} = \dfrac{\sqrt{1+\gamma^2}}{n_g}.
    \end{equation}
\end{itemize}

\section{Tested models}\label{sec:MG}
Future GWs observations are expected to contribute significantly  to probing  gravity~\citep{LISACosmologyWorkingGroup:2019}. Forecasts on the cross-correlation of the GWs signal with other probes suggest that the multi-messenger approach could be a powerful tool to exploit GWs observations to constrain models beyond $\Lambda$CDM~\citep{Mukherjee:gwxlss1,Mukherjee:gwxcmb,Balaudo22}. The GWs luminosity distance, for bright events, could provide a new probe to test gravity. In this work we discuss if future GWs observations combined with LSS probes could add new information on MG theories. We parametrize the effects of MG in a phenomenological way by adopting a general prescription suited to probe small departures from GR. In this section, we give a brief overview of the formalism we adopt and the models we investigate.

\subsection{Phenomenological parametrizations}
\begin{table}[]
    \centering
    \begin{tabular}{|c|c|c|c|c|c|c|}
    \hline
       Parameter & $\ln 10^{10}A_s$& $n_s$ & $w_0$ & $w_a$ & $E_{11}$ & $E_{22}$ \\ \hline
       Fiducial value & $3.098$ & $0.9619$ & $-1.00$ & $0.00$ & $0.18$  & $0.80$\\
         \hline
    \end{tabular}
    \caption{Assumed fiducial cosmology~\citep{planck:2018}. MG parameters are from Planck 2018 TT, TE, EE + lowE. Fiducial values for the $E_{ii}$ parameters lead to fiducial values on $\{\mu_0,\eta_0,\Sigma_0\}=\{1.12,1.55,1.43\}$.}
    \label{tab:fiducial_cosmology}
\end{table}
\begin{figure}[h]
	\centering
	\includegraphics[width=0.9\linewidth]{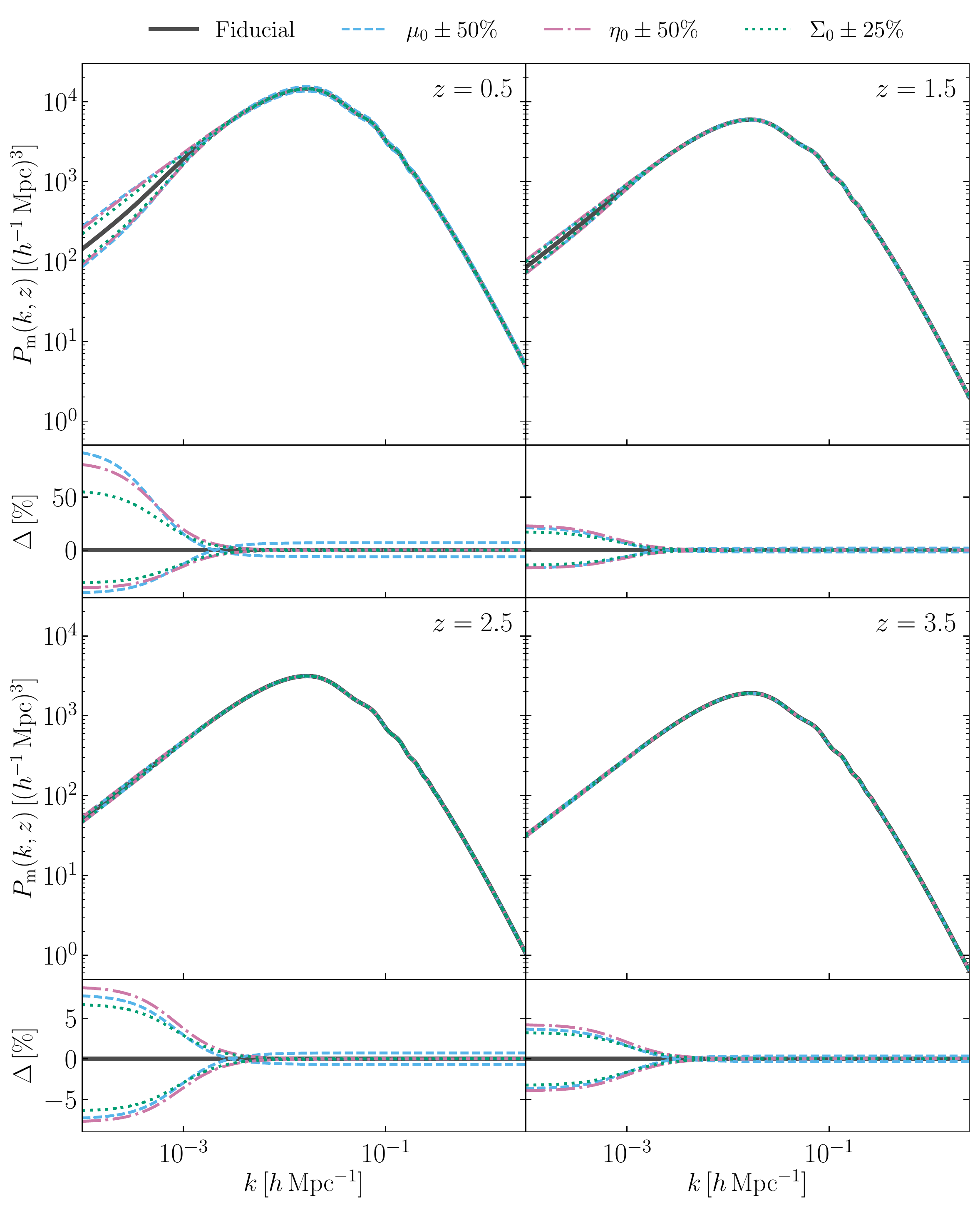}
	\caption{\textit{Upper panels}: predicted matter power spectrum for different values of the MG parameters at redshifts $z=\{0.5,\, 1.5,\, 2.5,\, 3.5\}$. We show $P_{\rm m}(k,z)$ for the assumed fiducial cosmology (solid black lines, see table~\ref{tab:fiducial_cosmology}) and for variations of $\mu_0$ (dashed light blue lines), $\eta_0$ (dashed-dotted pink lines) and $\Sigma_0$ (dotted green lines). When varying $\Sigma_0$, we keep $\mu_0$ fixed. \textit{Lower panels}: percentage variations with respect to the fiducial cosmology.}
	\label{fig:MG_matterpower}
\end{figure}
Starting from the LSS sector, we focus on scalar perturbations to the metric in the conformal Newtonian gauges, with the line element given by
\begin{equation}
    {\rm d}s^2 = a^2 \left[- (1 + 2\Psi){\rm d}\tau^2 + (1 - 2\Phi) {\rm d}x^2 \right],
\end{equation}
where $a$ is the scale factor, $\tau$ is the conformal time, and the time and scale-dependent functions $\Psi$ and $\Phi$ describe the scalar perturbations of the metric: the Newtonian potential and spatial curvature inhomogeneities, respectively. Modifications of gravity impact the growth of structure and the evolution of the gravitational potentials, see e.g.~\cite{Silvestri:2009hh,Joyce:2014kja}. Interestingly,  these effects, on linear scales, can be fully captured by two functions of time and scale, e.g.~\cite{Bertschinger:2006aw,Amendola:2007,Pogosian:2007sw,Zhao:2008,Zhang:2007} 
\begin{equation}\label{eq:potentials_1}
    k^{2} \Psi \equiv - 4 \pi G a^{2} \mu(k, z) \rho\Delta\,,
\end{equation}
and
\begin{equation}\label{eq:potentials_2}
    \Phi / \Psi \equiv \eta(k, z)\,,
\end{equation}
where $\rho\Delta=\rho_{\rm m}\Delta_{\rm m} + \rho_{\rm r}\Delta_{\rm r}$, i.e., the sum of the matter (m) and radiation (r) contributions.
One can also define the function $\Sigma(k, z)$, that quantifies modifications to the lensing potential, as 
\begin{equation}\label{eq:sigma_weyl_potential}
    k^{2}(\Phi+\Psi) \equiv -8 \pi G a^{2} \Sigma(k,z) \rho\Delta.
\end{equation}
The three phenomenological functions $\mu(k,z)$, $\eta(k,z)$ and $\Sigma(k,z)$ are not independent. One should consider two of them at the time, e.g., the pair $(\mu,\eta)$ or $(\mu,\Sigma)$. It is possible to express $\Sigma(k,z)$ as a function of $\mu(k,z)$ and $\eta(k,z)$ as
\begin{equation}
\label{eq:sigma}
    \Sigma(k,z)=\frac{\mu(k,z)}{2}(1+\eta(k, z)).
\end{equation}
Deviations from $\Lambda$CDM are encoded in   $(\mu(k,z),\, \eta(k,z))$ or $(\mu(k,z),\, \Sigma(k,z))$, with  the $\Lambda$CDM case corresponding to $\mu(k,z)=1$, $\eta(k,z)=1$, $\Sigma(k,z)=1$. To give a more intuitive interpretation of the physical meaning of the involved quantities, let us specify that the $\Sigma$ function acts on relativistic particles, affecting mainly the lensing observable, whereas $\mu$ controls gravity effects on massive particles, controlling the growth of matter perturbations and affecting clustering. Finally, $\eta$, usually referred to as the gravitational slip parameter, cannot be directly connected to a constraining observable as the previous two functions. However, given that it quantifies differences between the two gravitational potential, its behaviour may be indicative of a breaking of the equivalence principle.

 Several parametrizations of the phenomenological functions have been explored and constrained, see e.g.~\citep{Ishak:2018his} for a review on recent results. In this work, we follow the approach of the Planck 2015  paper on dark energy and modified gravity~\citep{planck:2015de-mg}. We choose a time-dependent only parametrization for the evolution of $\mu(k,z)$ and $\eta(k,z)$, the so-called late-time parametrization
\begin{equation}
\begin{aligned}
\mu(z) & = 1+E_{11} \Omega_{\mathrm{DE}}(z) \\
\eta(z) & = 1+E_{22} \Omega_{\mathrm{DE}}(z).
\end{aligned}
\end{equation}
The evolution is set by the value of the parameters $E_{11}$ and $E_{22}$, while the background is kept fixed. This choice of parametrization simplifies the analysis and allows a direct comparison with the results of~\citep{planck:2015de-mg}. But there are also good reasons for not expecting any scale-dependence of the model to show up within the range of scales covered by the data that we consider. In fact, in order to satisfy local tests of gravity, these theories need to have a working screening mechanism, which suppresses any deviation from GR through environmental effects. Well known examples are the Chameleon and Vainshtein mechanism, see e.g.~\cite{Brax:2021zjq}. In both cases, the requirements for a successful screening effectively pushes the characteristic length scale of the model either into the small, non-linear scales (Chameleon case) or to very large, horizon-size scales (Vainshtein case). Let us point out that even while not working with a specific model, there are some assumptions that we make at the basis of our choice of parametrization. One such assumption is that modifications of gravity are relevant at late times; in this sense, we are linking them possibly to the source of cosmic acceleration, but more broadly to tests of gravity with large scale structure. Or, said in other words, we aim for this parametrization to broadly represent Horndeski models of gravity with luminal speed of sound, which is the theoretical framework on which our analysis is built.

We consider both the $(\mu(z),\, \eta(z))$ and the $(\mu(z),\, \Sigma(z))$ pair. In the latter case, $\Sigma(z)$ as a function of $E_{11}$ and $E_{22}$ is computed using equation~\eqref{eq:sigma}. When performing the Fisher analysis, we vary $E_{11}$ and $E_{22}$ and derive the predicted constraints on the parameters $(\mu_0,\, \eta_0)$ and $(\mu_0,\, \Sigma_0)$, where $\mu_0 \equiv \mu(z=0)$, $\eta_0 \equiv \eta(z=0)$, $\Sigma_0 \equiv \Sigma(z=0)$. 

We compute the theoretical matter power spectrum with the code \texttt{MGCAMB}\footnote{See \url{https://github.com/sfu-cosmo/MGCAMB}.}~\cite{Zhao:2008,Hojjati:2011,Zucca:2019}, the modified version of the Einstein-Boltzmann solver \texttt{CAMB}\footnote{See \url{https://camb.info/}.}~\citep{Lewis:2000}, extended to study modified gravity models within the phenomenological parametrization framework. In figure~\ref{fig:MG_matterpower}, we show the linear matter spectrum for the assumed fiducial cosmology (see table~\ref{tab:fiducial_cosmology}) and for different values of the MG parameters $(\mu_0,\, \eta_0)$ and $(\mu_0,\, \Sigma_0)$. This is the power spectrum used to compute the $C_\ell$s introduced in section \ref{sec:formalism_Cls}. We observe that the most significant modifications occur at large scales. Varying the parameters $\eta_0$ or $\Sigma_0$ affects only the larger scales, while $\mu_0$ has an impact on smaller scales too. The modifications become milder at higher redshifts, according to how the parametrization we chose performs.

The late-time parametrization has been studied in the literature in several contexts~\citep{planck:2015de-mg,planck:2018,Casas:2017, DES:2018, DES:2021, DES:2022} and current data sets do not show a significant preference for models beyond $\Lambda$CDM. Recently, a non-parametric Bayesian reconstruction of $\mu,\Sigma$, along with the dark energy density, from all available LSS and CMB data was performed in~\cite{Pogosian:2021mcs,Raveri:2021dbu}; while the outcome is consistent with $\Lambda$CDM within $2\sigma$, some interesting features in $\Sigma$ were identified as an imprint of cosmological tensions. 

The phenomenological functions $\mu, \Sigma$ and $\eta$ parametrize modifications of the dynamics of perturbations within the scalar sector. When including GWs, one should consider that tensor perturbations are generally also affected by modifications of gravity.  For the observables of interest in this work, the effects of modified gravity on GWs propagating on the FLRW background, can be encoded in the difference between the electromagnetic luminosity distance $d_{\rm L}^{\rm EM}(z)$ and the GW one $d_{\rm L}^{\rm GW}(z)$. The phenomenological function $\Xi(z)$, defined as
\begin{equation}
    \Xi(z) \equiv \dfrac{d_{\rm L}^{\rm GW}(z)}{d_{\rm L}^{\rm EM}(z)},
\end{equation}
quantifies the effect for bright sirens. The EM luminosity distance can be expressed as $d_{\rm L}^{\rm EM}(z)=\sqrt{L/4\pi S}$, where L and S are the bolometric luminosity and the bolometric flux for the observed object, respectively. This quantity can also be expressed as a function of the comoving distance $\chi$ as (for $\Omega_k=0$): $d_{\rm L}^{\rm EM}(z) = (1+z)\chi = (1+z)d_H\int_{0}^{z}dz'/E(z')$, with $E(z) = \sqrt{\Omega_m(1+z)^3+\Omega_k(1+z)^2+\Omega_\Lambda}$ and $d_H=c/H_0$. The GW luminosity distance $d_{\rm L}^{\rm GW}(z)$ is estimated in a way not dependent on a distance ladder, and relies on the extraction of information enclosed in the GW waveform such as the strain and the frequency. A univocal analytic expression for the $d^{GW}_L(z)$ is non trivial to obtain, as it is also highly dependent on the assumed gravity model. In~\citep{LISACosmologyWorkingGroup:2019}, the authors performed an extensive study of $\Xi$ both in terms of parametrizations and of specific form it takes in given models of MG. In the latter case, $\Xi$ is in general related to operators of the Lagrangian that affect also  scalar perturbations; for instance, in Horndeski gravity it is a function of the non-minimal coupling, which is a key contributor to $\mu$ and $\Sigma$ as well. Therefore, in a  theoretical embedding, $\Xi$ is not completely independent of $\mu$ and $\Sigma$. 

The expressions we used for GW lensing in the auto- and cross-correlation rely on calculations of the relativistic corrections to the luminosity distance of GW in Horndeski and DHOST theories with the speed of sound $c_T^2=1$~\cite{Tasinato:2021wol}. In this case, it is not straightforward to find an explicit expression for $\Xi(z)$ in terms of $\mu,\Sigma$ and/or $\eta$ which is valid on all linear scales. For $c_T^2=1$,   in the quasi-static regime and on scales above the mass scale of the model, the running of the Planck mass is the main contributor to both $\mu$,  implying that the relation between $\Xi$ and $\mu$ tends towards the simple form 
\begin{align}
    \begin{split}
        \Xi(z) & = \sqrt{1 + \frac{1}{\mu (z)}}.
    \end{split}
\end{align}
However, on smaller scales, the relation becomes more complicated, as discussed in~\cite{Pogosian:2016pwr}, and the expression for  $\Xi$ would acquire another term, dependent on the other MG functions at play. For the parametrization of $\mu$ and $\eta$ that we employ in this work, based on~\cite{planck:2015de-mg}, the exact form of this additional term, which should depend on $\eta$, is complex to work out without loosing generality. For this reason, we decide to parametrize the $\Xi$ function as follows
\begin{align}\label{eq:Xi_a1a2}
    \begin{split}
        \Xi(z) & = \sqrt{1 + \frac{1}{\mu (z)} + \frac{a_1}{\eta(z)^{a_2}}},
    \end{split}
\end{align}
where $a_1$ and $a_2$ are varied along with the other parameters in the Fisher analysis and regarded as nuisances. With this parametrization of $\Xi$ we can reproduce the main features of the results found for several models~\cite{LISACosmologyWorkingGroup:2019}. We fix the fiducial values of $a_1$ and $a_2$ in order to obtain a variation of $\Xi$ in redshift comparable to the results for DHOST models in~\cite{LISACosmologyWorkingGroup:2019}.

Let us stress that our method for GW lensing builds on the expressions for the luminosity distance of GWs and its relativistic corrections; the latter are explicitly known only for the class of Horndeski models with luminal speed of tensors. This is therefore the context in which we perform our analysis. In this framework, the $\Xi$ function is not independent of the $\mu, \eta$ or $\mu, \Sigma$ functions. In other words, they all depend, solely or partially, on the non-minimal coupling of the theory. A more general framework may not encode this dependence; forecasts in such case would be expected to be less constraining. In order to correctly quantify the degrading, we would need to go beyond the theoretical framework on which we have built our analysis; this is certainly an interesting direction for future work.

In the following section we discuss how the observables that we consider in this work are modified in light of the MG phenomenological functions.  
\subsection{Angular power spectra (in MG) and MG parameters}
Above, we commented on how the MG parameters affect the linear matter power spectrum (see figure~\ref{fig:MG_matterpower}). In this section, we outline their impact on the observables that we consider in this work, presented in section~\ref{sec:formalism_Cls}.

On the one hand, all the angular power spectra are computed with the linear matter power spectrum $P_{\rm mm}(k,z)$. In our analysis the modified $P_{\rm mm}(k,z)$ is computed numerically by means of the code \texttt{MGCAMB}, as discussed above. As can be noticed in figure \ref{fig:MG_matterpower}, the MG functions affect the matter power spectrum $P_{\rm mm}(k)$. The effect of $\mu$ is quite direct and the most notable, given that $\mu$ changes the rate of clustering of matter. The functions $\eta$ and $\Sigma$ have a less direct impact on $P_{\rm mm}(k)$, but still affect it. In particular, $\Sigma$ impacts the C spectrum via the magnification bias. On the other hand, the MG models we consider modify the lensing potential. In the scalar sector, modifications to the lensing potential are encoded by the MG function $\Sigma(z)$, through equation~\eqref{eq:sigma_weyl_potential}. This means that an extra factor $\Sigma(z)$ appears for each time the term $(\Phi + \Psi)$ appears. Thus, the lensing angular power spectra of equation~\eqref{eq:weak_lensing_LL} become \cite{Casas:2017}
\begin{equation}
    C_\mathrm{LL}^{\rm X_iY_j}(\ell) =
    \int_0^\infty \frac{d z \ c}{H(z)} \  \frac{W_\mathrm L^{\rm X_i}(z) \ W_\mathrm L^{\rm Y_j}(z)}{\chi^2(z)} \ \Sigma^2(z)\  P_\mathrm{mm}\left(\frac{\ell}{\chi(z)},z\right),
    \label{eq:weak_lensing_LL_MG}
\end{equation}
 while the cross-correlation spectra between lensing and clustering will be
\begin{equation}
    C_\mathrm{CL}^{\rm X_iY_j}(\ell) = \int_0^\infty \frac{d z \ c}{H(z)} \ \frac{W_\mathrm C^{\rm X_i}\left(\frac{\ell}{\chi(z)},z\right) \ W_\mathrm L^{\rm Y_j}(z)}{\chi^2(z)} \ \Sigma(z) P_\mathrm{mm}\left(\frac{\ell}{\chi(z)},z\right).
    \label{eq:weak_lensing_GL_MG}
\end{equation}
Following \cite{Balaudo22}, the $\Xi(z)$ function is going to appear in the Lensing observables related to bright GW sirens. This is because for bright sirens the estimator of the convergence depends on the ratio $d_{\rm L}^{\rm GW}/d_{\rm L}^{\rm EM}$. Explicitly, this results into
\begin{equation}
    C_\mathrm{LL}^{\rm GW^{bright}_i GW^{bright}_j}(\ell) \simeq \Xi^2(z)
    \int_0^\infty \frac{d z' \ c}{H(z')} \  \frac{W_\mathrm L^{\rm X_i}(z') \ W_\mathrm L^{\rm Y_j}(z')}{\chi^2(z')} \ \Sigma^2(z')\  P_\mathrm{mm}\left(\frac{\ell}{\chi(z')},z'\right),
\end{equation}
 while the cross-correlation spectra between lensing and clustering will be
\begin{equation}
    C_\mathrm{L\Theta}^{\rm GW^{bright}_i Y_j}(\ell) \simeq \Xi(z) \int_0^\infty \frac{d z' \ c}{H(z')} \ \frac{W_\mathrm L^{^{\rm GW^{bright}_i Y_j}}\left(\frac{\ell}{\chi(z')},z'\right) \ W_\mathrm L^{\rm Y_j}(z')}{\chi^2(z')} \ \Sigma(z') P_\mathrm{mm}\left(\frac{\ell}{\chi(z')},z'\right).
\end{equation}
Following reference \cite{Balaudo22}, it is worth clarifying that the approximately equal symbol in the above two equations is due to the linearization at first order of the convergence estimator, in the parameters describing it which are introduced in equation (3.8) of \cite{Balaudo22}. We refer the interested reader to reference \cite{Balaudo22} for further details.

\section{Forecasts}\label{sec:forecasts}
In accordance to what described in section \ref{sec:formalism_Fisher}, we perform a Fisher analysis on the following parameters: $\{E_{11}, E_{22}, w_0, w_a, \ln 10^{10}A_s, n_s, K^{\mathrm{fg}},a_1,a_2\}$ (for a total of 9 parameters). The fiducial values we use in this pipeline (mainly taken from Planck results \cite{planck:2018}) are summarized in table~\ref{tab:fiducial_cosmology}.\footnote{The fiducial value of $K^{\mathrm{fg}}$ depends on the case considered (z binning and probe): we adopt $K^{\mathrm{fg}}=5.72 \cdot 10^{-8}$ ($9.64 \cdot 10^{-6}$) for the redshift binning chosen in the dark sirens case for the Lensing (Clustering) probe and $K^{\mathrm{fg}}=4.49 \cdot 10^{-8}$ ($1.05 \cdot 10^{-5}$) for the redshift binning chosen in the bright sirens case for the Lensing (Clustering) probe. The fiducial values for \{$a_1,a_2$\} are respectively -0.95 and 0.14. We omit these values from table~\ref{tab:fiducial_cosmology} for the sake of simplicity.} Where explicitly stated, the $\{w_0, w_a\}$ parameters are kept fixed instead. Given errors on $\{E_{11}, E_{22}\}$, we derive constraints on the $\{\mu_0, \eta_0, \Sigma_0\}$ parameters. We perform Fisher analysis for the following different cases:
\begin{itemize}
    \item Different probes: L only, C only, and $\mathrm{L + C}$;
    \item Different tracers combinations: $\mathrm{GW \times IM}$ and $\mathrm{GW \times IM \times gal}$;
    \item GW are either treated as dark (BHBH and BHNS mergers) or bright (NSNS) sirens.
\end{itemize}
The next section provides results on $\{\mu_0, \Sigma_0\}$ for all the cases listed above. Appendix \ref{app:mueta} provides the same results for $\{\mu_0, \eta_0\}$.

\subsection{Results}
We provide Fisher estimated constraints on the \{$\mu_0, \eta_0, \Sigma_0\}$ parameters (and $w_0,w_a$ where relevant) in tables \ref{tab:errors_darkbirght_withw0wa} and \ref{tab:errors_darkbirght_NOw0wa}. All results refer to $f_{\rm sky}=0.5$ and $T_{\rm obs}^{\rm GW}=15$yr.

\begin{table}[]
\centering
\begin{tabular}{l|ccccc|ccccc|}
\cline{2-11}
                                 & \multicolumn{1}{c|}{$\sigma_{\mu_0}$} & \multicolumn{1}{c|}{$\sigma_{\eta_0}$} & \multicolumn{1}{c|}{$\sigma_{\Sigma_0}$} & \multicolumn{1}{c|}{$\sigma_{w_0}$} & $\sigma_{w_a}$                     & \multicolumn{1}{c|}{$\sigma_{\mu_0}$} & \multicolumn{1}{c|}{$\sigma_{\eta_0}$} & \multicolumn{1}{c|}{$\sigma_{\Sigma_0}$} & \multicolumn{1}{c|}{$\sigma_{w_0}$} & {$\sigma_{w_a}$}                      \\ \cline{2-11} 
                                 & \multicolumn{5}{c|}{\textbf{darkGW$\times$IM}}                                                                                                                                      & \multicolumn{5}{c|}{\textbf{brightGW$\times$IM}}                                                                                                                                     \\ \hline
\multicolumn{1}{|l|}{LENSING}    & \multicolumn{1}{c|}{15.62}          & \multicolumn{1}{c|}{38.92}           & \multicolumn{1}{c|}{2.09}              & \multicolumn{1}{c|}{2.78}  & 8.66                     & \multicolumn{1}{c|}{24.45}           & \multicolumn{1}{c|}{61.86}            & \multicolumn{1}{c|}{3.73}              & \multicolumn{1}{l|}{4.59}  & \multicolumn{1}{l|}{16.18} \\ \hline
\multicolumn{1}{|l|}{CLUSTERING} & \multicolumn{1}{c|}{1.20}           & \multicolumn{1}{c|}{1.84}            & \multicolumn{1}{c|}{0.99}              & \multicolumn{1}{c|}{0.53}  & 1.43                      & \multicolumn{1}{c|}{1.12}           & \multicolumn{1}{c|}{1.84}            & \multicolumn{1}{c|}{1.00}              & \multicolumn{1}{c|}{0.47}  & 1.34                       \\ \hline
\multicolumn{1}{|l|}{L + C}      & \multicolumn{1}{c|}{0.10}           & \multicolumn{1}{c|}{0.24}            & \multicolumn{1}{c|}{0.04}              & \multicolumn{1}{c|}{0.11}  & 0.23                     & \multicolumn{1}{c|}{0.08}           & \multicolumn{1}{c|}{0.20}            & \multicolumn{1}{c|}{0.06}              & \multicolumn{1}{c|}{0.05}  & 0.15                       \\ \hline
                                 & \multicolumn{5}{c|}{\textbf{darkGW$\times$IM$\times$gal}}                                                                                                                                  & \multicolumn{5}{c|}{\textbf{brightGW$\times$IM$\times$gal}}                                                                                                                                 \\ \hline
\multicolumn{1}{|l|}{LENSING}    & \multicolumn{1}{c|}{1.96}           & \multicolumn{1}{c|}{4.48}            & \multicolumn{1}{c|}{0.09}              & \multicolumn{1}{c|}{0.45}  & 1.41                     & \multicolumn{1}{c|}{2.78}           & \multicolumn{1}{c|}{6.66}            & \multicolumn{1}{c|}{0.24}              & \multicolumn{1}{c|}{0.96}  &   3.62                     \\ \hline
\multicolumn{1}{|l|}{CLUSTERING} & \multicolumn{1}{c|}{0.80}           & \multicolumn{1}{c|}{1.39}            & \multicolumn{1}{c|}{0.82}              & \multicolumn{1}{c|}{0.32}  & 0.92                      & \multicolumn{1}{c|}{0.30}           & \multicolumn{1}{c|}{1.25}            & \multicolumn{1}{c|}{0.68}              & \multicolumn{1}{c|}{0.12}  & 0.35                       \\ \hline
\multicolumn{1}{|l|}{L + C}      & \multicolumn{1}{c|}{0.05}           & \multicolumn{1}{c|}{0.11}            & \multicolumn{1}{c|}{0.01}              & \multicolumn{1}{l|}{0.04}  & \multicolumn{1}{l|}{0.08} & \multicolumn{1}{c|}{0.06}           & \multicolumn{1}{c|}{0.10}            & \multicolumn{1}{c|}{0.02}              & \multicolumn{1}{c|}{0.03}  & 0.09                       \\ \hline
\end{tabular}
\caption{Fisher estimated errors on the $\mu_0, \eta_0, \Sigma_0$, $w_0, w_a$ parameters for different tracers and probes combinations.}
\label{tab:errors_darkbirght_withw0wa}
\end{table}

\begin{table}[h!]
\centering
\begin{tabular}{l|ccc|ccc|}
\cline{2-7}
                                 & \multicolumn{1}{c|}{$\sigma_{\mu_0}$} & \multicolumn{1}{c|}{$\sigma_{\eta_0}$} & $\sigma_{\Sigma_0}$ & \multicolumn{1}{c|}{$\sigma_{\mu_0}$} & \multicolumn{1}{c|}{$\sigma_{\eta_0}$} & $\sigma_{\Sigma_0}$ \\ \cline{2-7} 
                                 & \multicolumn{3}{c|}{\textbf{darkGW$\times$IM}}                                                        & \multicolumn{3}{c|}{\textbf{brightGW$\times$IM}}                                                      \\ \hline
\multicolumn{1}{|l|}{LENSING}    & \multicolumn{1}{c|}{10.46}          & \multicolumn{1}{c|}{25.46}           & 1.06              & \multicolumn{1}{c|}{16.38}           & \multicolumn{1}{c|}{40.61}            & 2.06              \\ \hline
\multicolumn{1}{|l|}{CLUSTERING} & \multicolumn{1}{c|}{0.18}           & \multicolumn{1}{c|}{1.10}            & 0.62              & \multicolumn{1}{c|}{0.19}           & \multicolumn{1}{c|}{1.23}            & 0.68              \\ \hline
\multicolumn{1}{|l|}{L + C}      & \multicolumn{1}{c|}{0.09}           & \multicolumn{1}{c|}{0.20}            & 0.02              & \multicolumn{1}{c|}{0.03}           & \multicolumn{1}{c|}{0.09}            & 0.03              \\ \hline
                                 & \multicolumn{3}{c|}{\textbf{darkGW$\times$IM$\times$gal}}                                                    & \multicolumn{3}{c|}{\textbf{brightGW$\times$IM$\times$gal}}                                                  \\ \hline
\multicolumn{1}{|l|}{LENSING}    & \multicolumn{1}{c|}{1.13}           & \multicolumn{1}{c|}{2.60}            & 0.06              & \multicolumn{1}{c|}{1.63}           & \multicolumn{1}{c|}{3.93}            &  0.14             \\ \hline
\multicolumn{1}{|l|}{CLUSTERING} & \multicolumn{1}{c|}{0.17}           & \multicolumn{1}{c|}{1.08}            & 0.61              & \multicolumn{1}{c|}{0.09}           & \multicolumn{1}{c|}{0.50}            & 0.29              \\ \hline
\multicolumn{1}{|l|}{L + C}      & \multicolumn{1}{c|}{0.08}           & \multicolumn{1}{c|}{0.17}            & 0.02              & \multicolumn{1}{c|}{0.03}           & \multicolumn{1}{c|}{0.06}            & 0.01              \\ \hline
\end{tabular}
\caption{Fisher estimated errors on the $\mu_0, \eta_0, \Sigma_0$ parameters for different tracers and probes combinations. The parameters $w_0, w_a$ are kept fixed.}
\label{tab:errors_darkbirght_NOw0wa}
\end{table}

\begin{figure}[h!]
	\centering
	\includegraphics[width=0.75\linewidth]{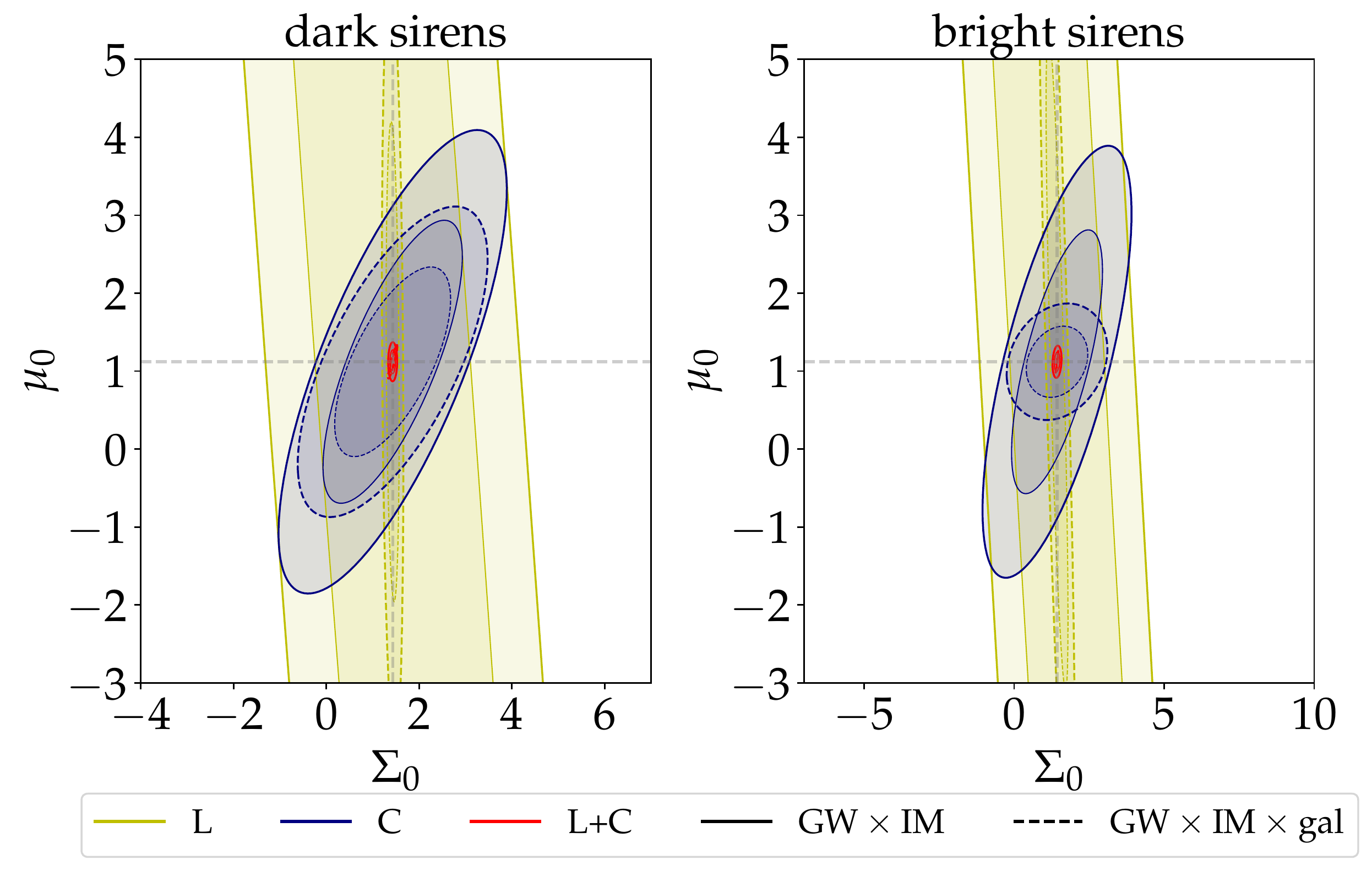}
	\caption{Contours for $\{\mu_0,\Sigma_0\}$ in the GW$\times$IM (solid line) and GW$\times$IM$\times$gal (dashed line) cases, all probes considered (colors according to legend). Left panels refer to dark sirens (BHBH+BHNS), right panels refer to bright sirens (NSNS). $w_0, w_a$ are among the Fisher parameters considered. $T_{\rm obs}^{\rm GW}$=15 yr and $f_{\rm sky}=0.5$.}
	\label{fig:musigma_ellipses_w0wacase}
\end{figure}

\begin{figure}
	\centering
	\includegraphics[width=0.78\linewidth]{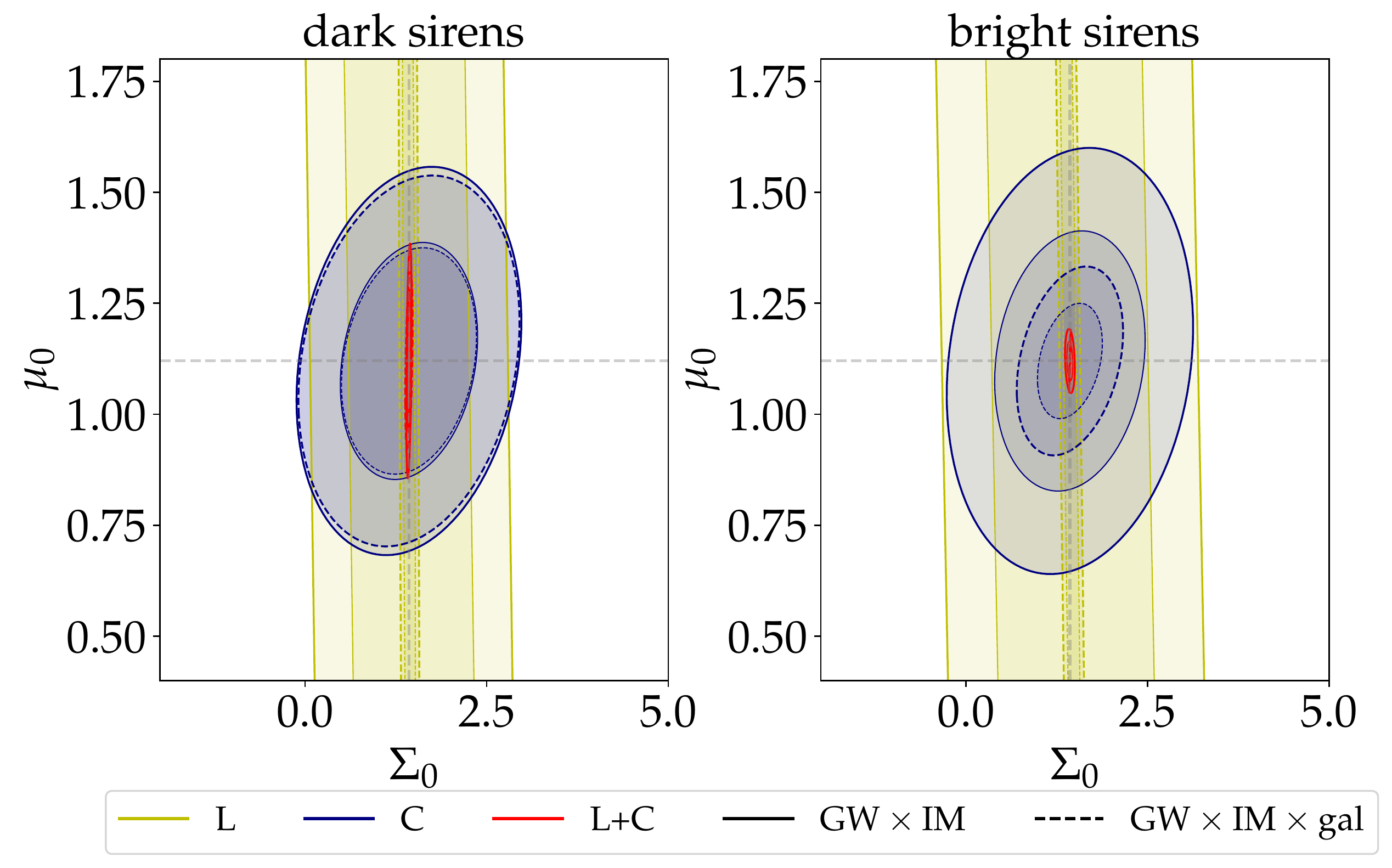}
	\caption{Contours for $\{\mu_0,\Sigma_0\}$ in the GW$\times$IM (solid line) and GW$\times$IM$\times$gal (dashed line) cases, all probes considered (colors according to legend). Left panels refer to dark sirens (BHBH+BHNS), right panels refer to bright sirens (NSNS). $w_0, w_a$ are fixed to fiducial values. $T_{\rm obs}^{\rm GW}$=15 yr and $f_{\rm sky}=0.5$.}
	\label{fig:musigma_ellipses_NOw0wacase}
\end{figure}

\begin{figure}
	\centering
	\includegraphics[width=0.8\linewidth]{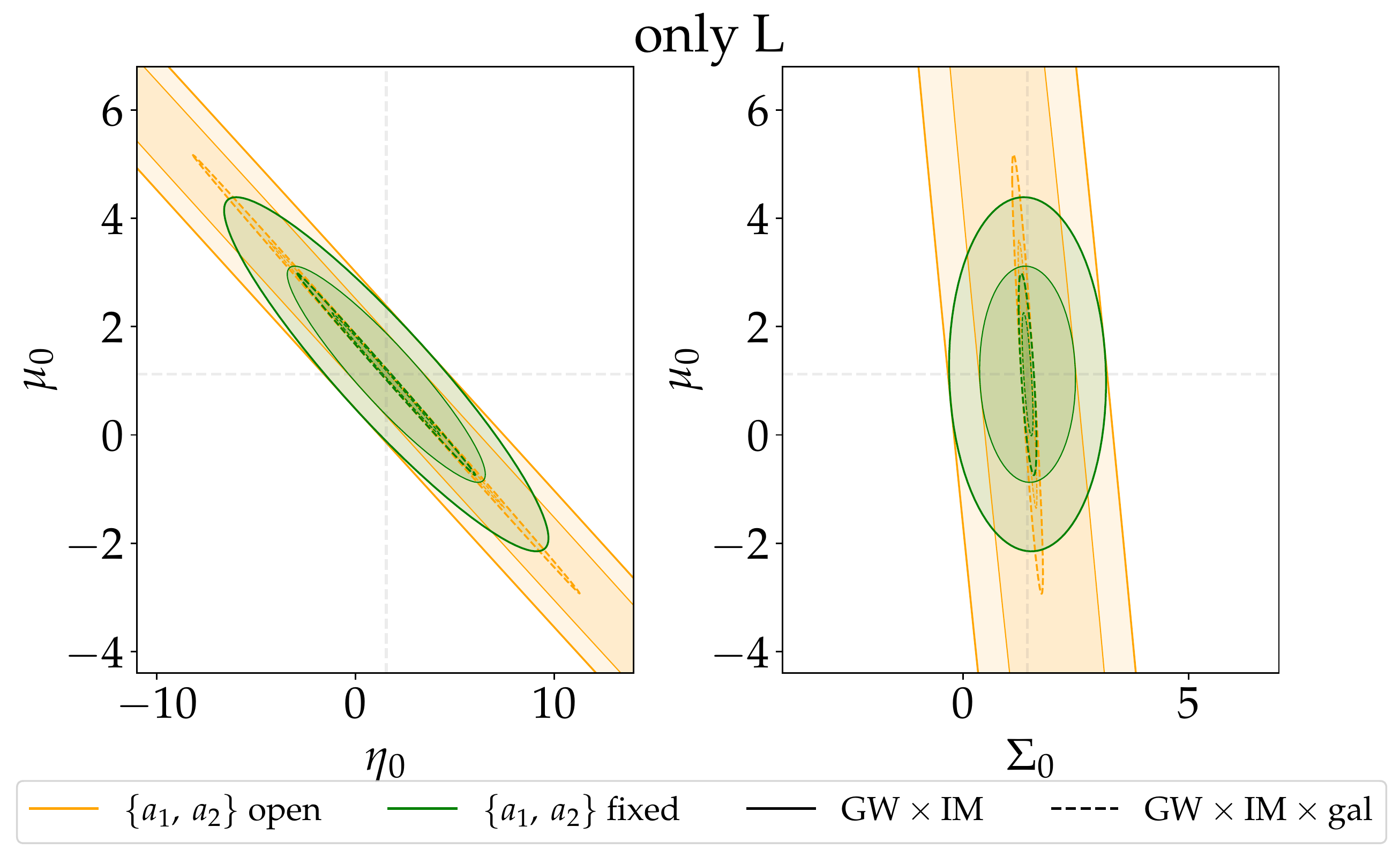}
	\caption{Contours for $\{\mu_0,\eta_0\}$ and $(\mu_0,\Sigma_0)$ in the GW$\times$IM (solid line) and GW$\times$IM$\times$gal (dashed line) cases, for the only-L case. Comparison between forecasts fixing or opening the $\{a_1,a_2\}$ parameters describing the $\Xi$ function according to equation \eqref{eq:Xi_a1a2}. $w_0, w_a$ are fixed to fiducial values. $T_{\rm obs}^{\rm GW}$=15 yr and $f_{\rm sky}=0.5$.}
	\label{fig:ellipses_Lonly}
\end{figure}

\begin{figure}
	\centering
	\includegraphics[width=1.0\linewidth]{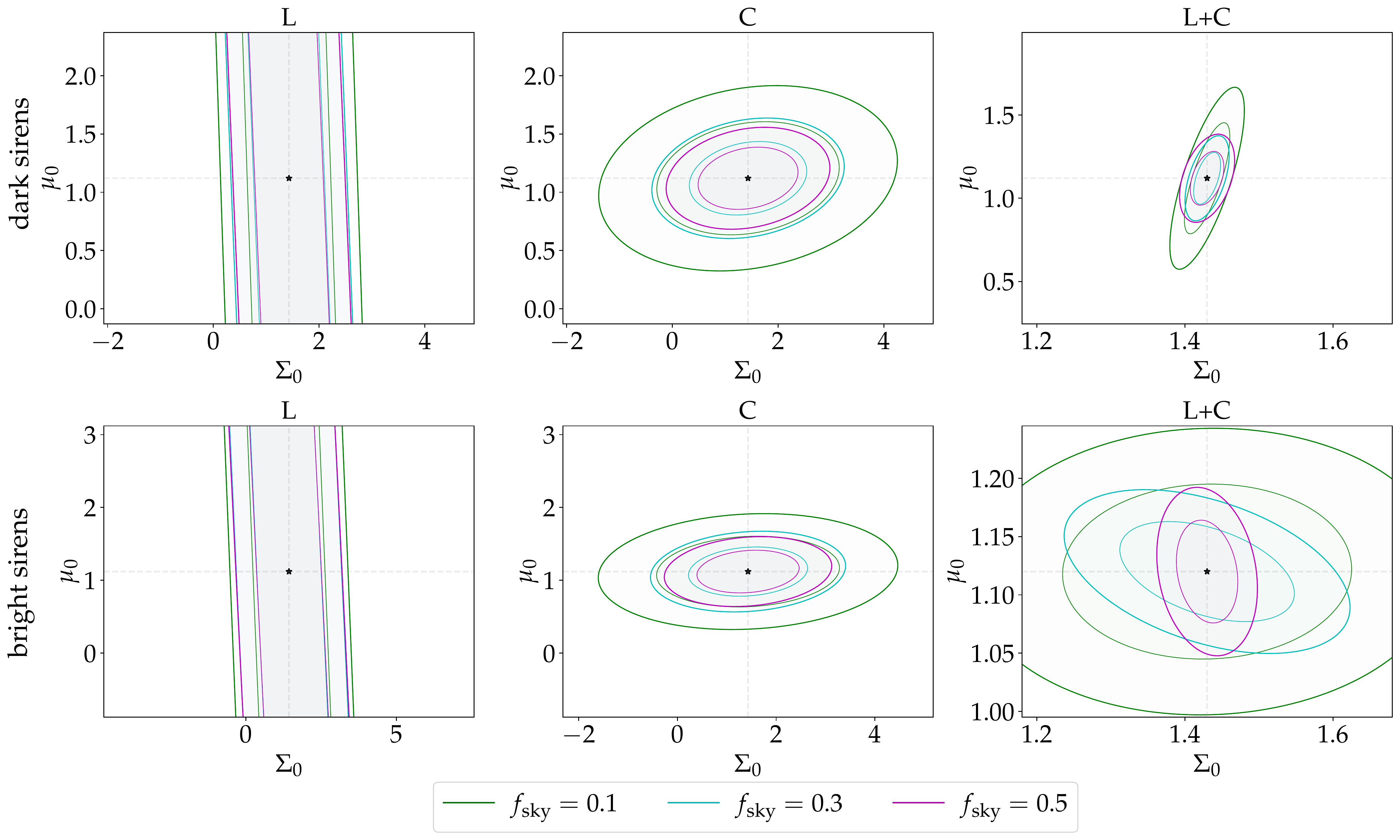}
	\caption{Contours for $\{\mu_0,\Sigma_0\}$ parameters for the GW$\times$IM case, all probes considered (only Lensing: left panels; only Clustering: center panels; Lensing + Clustering: right panels), for dark (top panels) and bright (bottom panels) sirens and for different values of $f_{\rm sky}$ (colour-coded according to legend). $w_0, w_a$ are fixed to fiducial values. $T_{\rm obs}^{\rm GW}$=15 yr.}
	\label{fig:muSigma_ellipses_fsky_GWxIM_NOw0wacase}
\end{figure}

\begin{figure}
	\centering
	\includegraphics[width=1.0\linewidth]{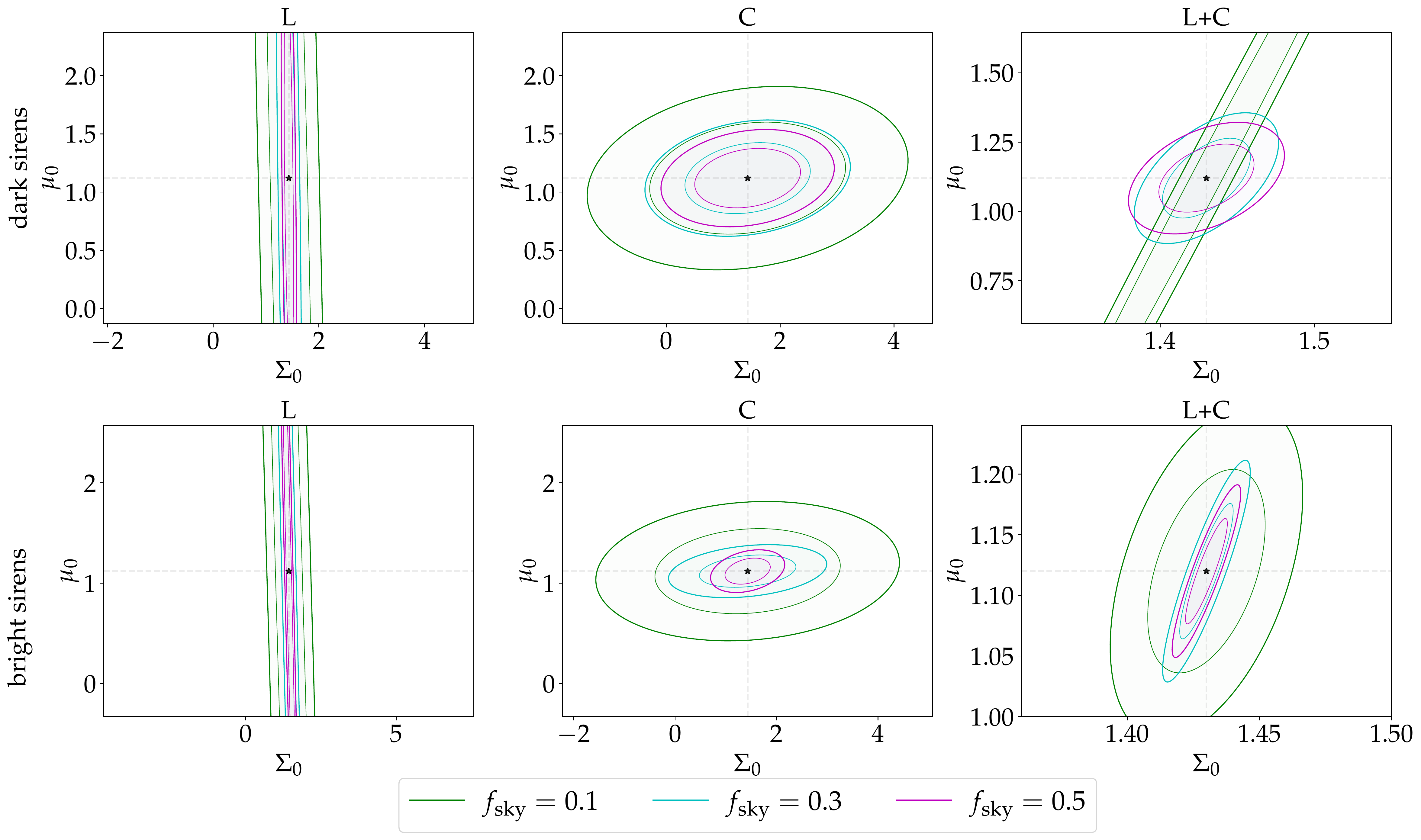}
	\caption{Contours for $\{\mu_0,\Sigma_0\}$ parameters for the GW$\times$IM$\times$gal case, all probes considered (only Lensing: left panels; only Clustering: center panels; Lensing + Clustering: right panels), for dark (top panels) and bright (bottom panels) sirens and for different values of $f_{\rm sky}$ (colour-coded according to legend). $w_0, w_a$ are fixed to fiducial values. $T_{\rm obs}^{\rm GW}$=15 yr.}
	\label{fig:muSigma_ellipses_fsky_GWxIMxgal_NOw0wacase}
\end{figure}

\begin{figure}
	\centering
	\includegraphics[width=0.95\linewidth]{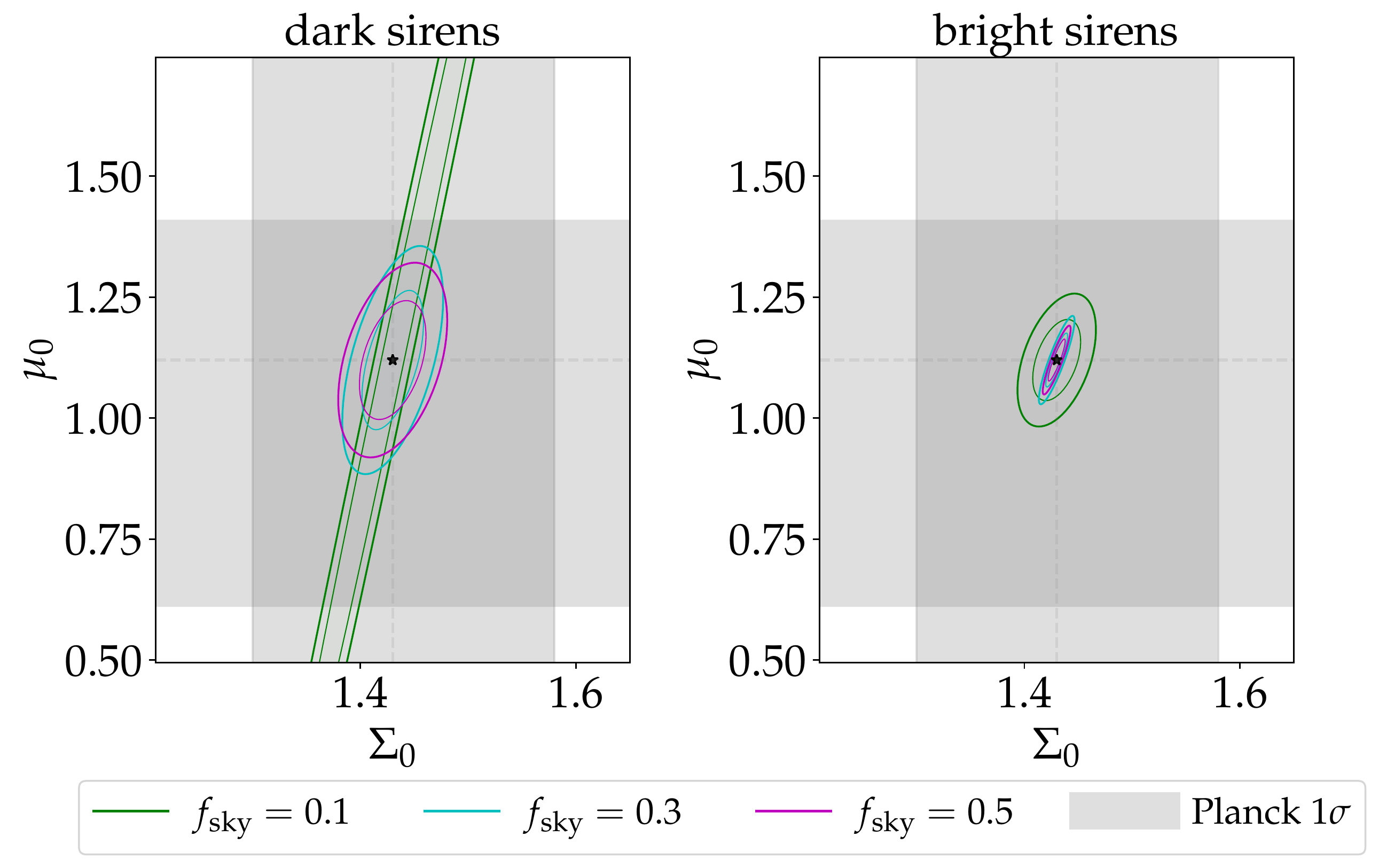}
	\caption{Contours for the most stringent constraints we find on $\{\mu_0,\Sigma_0\}$: GW$\times$IM$\times$gal, Lensing + Clustering case, for dark (left) and bright (right) sirens and for different values of $f_{\rm sky}$ (colour-coded according to legend). $w_0, w_a$ are fixed to fiducial values. $T_{\rm obs}^{\rm GW}$=15 yr. The gray area shows $1\sigma=68\%$ confidence regions from Planck TT,TE,EE+lowE without CMB lensing (see table 7 of \cite{planck:2018}).}
	\label{fig:muSigma_ellipses_fsky_GWxIMxgal_NOw0wacase_Planck}
\end{figure}

In figures \ref{fig:musigma_ellipses_w0wacase}, \ref{fig:musigma_ellipses_NOw0wacase} we provide $1-2 \sigma$ contour ellipses on the \{$\mu_0, \Sigma_0\}$ parameters (for $f_{\rm sky}=0.5$ and $T_{\rm obs}^{\rm GW}=15$yr). In figures \ref{fig:muSigma_ellipses_fsky_GWxIM_NOw0wacase},  \ref{fig:muSigma_ellipses_fsky_GWxIMxgal_NOw0wacase} we provide forecasts on $\{\mu_0, \Sigma_0\}$ for different values of $f_{\rm sky}$, fixing $w_0, w_a$. The same plots for the \{$\mu_0, \eta_0\}$ parameters are provided in appendix \ref{app:mueta} (figures \ref{fig:mueta_ellipses_w0wacase}, \ref{fig:mueta_ellipses_NOw0wacase}, \ref{fig:mueta_ellipses_fsky_GWxIM_NOw0wacase}, \ref{fig:mueta_ellipses_fsky_GWxIMxgal_NOw0wacase}). In light of these results, we can express the statements in the following.

\subsubsection*{\textit{Lensing-only case}}\label{sec:forecasts_Lonly}
Focusing on the Lensing-only case, we find that both bright and dark sirens cases are not good at constraining the parameters of interest, although some differences in the constraining power between the two cases can be found. Indeed, considering bright sources brings both advantages and disadvantages, with the resulting outcome depending on which of the two dominates. Specifically, the advantage of having an EM counterpart is enclosed in the presence of the MG function $\Xi$ defined in section \ref{sec:MG} (not present for dark sirens), which introduces a severely stronger dependence of the $C_{\ell}$s on the $\mu_0, \eta_0, \Sigma_0$ parameters. On the other side, detectable bright sources cover a lower redshift range (since NSNS binaries are less massive than BHBH or BHNS they can be detected up to lower redshifts). This might give a disadvantage, both concerning the number of detected sources (i.e., worse shot noise) and the possibility to perform a less deep tomography (fewer redshift bins available, i.e., less information). Overall, bright sirens may give better/worse results with respect to the dark case depending on the balance between these two effects and on which probe we are considering. 

Generally, in the L-only case the advantages of considering bright sirens are not able to dominate on the downsides (or at least significantly), with constraints comparable between the two cases (see e.g., tables \ref{tab:errors_darkbirght_withw0wa} and \ref{tab:errors_darkbirght_NOw0wa}). This is even more evident in e.g., figures \ref{fig:musigma_ellipses_w0wacase}, \ref{fig:musigma_ellipses_NOw0wacase}, \ref{fig:mueta_ellipses_w0wacase}, \ref{fig:mueta_ellipses_NOw0wacase}: the L-only contour ellipses (in yellow) show an extremely wide extension in all dark sirens panels (left side), leaving especially $\mu_0$ and $\eta_0$ barely constrained. Unfortunately, a similar trend can be found for bright sirens L-only ellipses (right panels of both figures). The same insight can be drawn from figures \ref{fig:muSigma_ellipses_fsky_GWxIM_NOw0wacase}, \ref{fig:muSigma_ellipses_fsky_GWxIMxgal_NOw0wacase}, \ref{fig:mueta_ellipses_fsky_GWxIM_NOw0wacase}, \ref{fig:mueta_ellipses_fsky_GWxIMxgal_NOw0wacase}.

Furthermore, we can see that adding galaxies in addition to the $\mathrm{GW \times IM}$ cross-correlation significantly improves the results, especially in the dark sirens case: dashed lines (GW$\times$IM$\times$gal) in figures \ref{fig:musigma_ellipses_w0wacase} and \ref{fig:musigma_ellipses_NOw0wacase} tend to mark tighter ellipses than solid lines (GW$\times$IM).

Overall, Lensing-only forecasts are non-competitive with Planck constraints~\cite{planck:2018}, showing nonetheless the advantage of taking into account the information coming from a higher number of tracers (GW$\times$IM$\times$gal vs. GW$\times$IM).

\subsubsection*{\textit{L+C case}}
Adding the angular Clustering probe to Lensing data (L+C case) significantly improves the results in any case considered (bright/dark sirens, with/without adding resolved galaxies), providing constraints tighter up to two orders of magnitude (see e.g., table \ref{tab:errors_darkbirght_NOw0wa}). This shows that not only cross-correlating different tracers but especially combining together different probes is a remarkably powerful tool to exploit, that provides significant extra information. This is especially evident in figures \ref{fig:musigma_ellipses_w0wacase} and \ref{fig:mueta_ellipses_w0wacase}: the C-only (in blue) and especially the L+C (in red) contours are firmly more constraining than the (yellow) L-only ones, often breaking down degeneracies between parameters.

The best results we obtain in the L+C case are very competitive with Planck results~\cite{planck:2018}, highlighting the power of cross-combining observables of different tracers and probes. Results concerning the $\Sigma_0$ parameter are especially promising. This is reasonable since $\Sigma_0$ is the parameter describing deviations from GR for Lensing effects, as explained in section \ref{sec:MG}. To highlight the competitiveness of our best constraints with those from Planck, in figure \ref{fig:muSigma_ellipses_fsky_GWxIMxgal_NOw0wacase_Planck} we compare our L+C forecasts (GW$\times$IM$\times$gal case) with the $68\%$ confidence regions from Planck TT,TE,EE+lowE (without CMB lensing, see table 7 of [127]). Planck results are compatible with $\Lambda$CDM and Planck data alone do not show a significant preference for beyond $\Lambda$CDM values of $\mu_0$, $\eta_0$ and $\Sigma_0$: indeed, their results are less than $1\sigma$ away from the $\Lambda$CDM limit for $\mu_0$ and $\eta_0$, and $\sim 2\sigma$ for $\Sigma_0$. Our best results are highly competitive and severely reduce Planck errors: assuming a Planck best fit as fiducial value our measurements show a mild preference for non-$\Lambda$CDM values of  $\mu_0$ and $\eta_0$ (respectively $\sim4 \sigma$ and $\sim9 \sigma$), and a clearly stronger preference for $\Sigma_0$ (at more than 20$\sigma$), since our lensing observable is strongly affected by it. This means that if experimental data will confirm beyond $\Lambda$CDM central values, we would be able to confirm a preference for MG models with a high confidence level.

Comparing the bright/dark sirens cases, we see no univocal pattern among the two (see e.g., red L+C contours in figures \ref{fig:musigma_ellipses_w0wacase} and \ref{fig:musigma_ellipses_NOw0wacase}) . This can be motivated by the explanation laid in the previous point: taking bright sirens has both pros (extra information contained in the $\Xi$ parameter for Lensing) and cons (shallower tomography in both L and C). Given the addition of Clustering (which is independent of $\Xi$) we can not naturally expect a striking difference as for the L-only case, but a competition between these two opposite effects, with not clearly predictable outcomes. We also note that generally adding galaxies improves the constraining power, which is an expected outcome as more information is being fed to the pipeline (as for the L-only case).

\subsubsection*{\textit{Fixing $\{a_1,a_2\}$ parameters}}
In section \ref{sec:MG} we have introduced the $\Xi(z)$ function, which is parametrized by $\{a_1,a_2\}$ according to equation \eqref{eq:Xi_a1a2}. In order to take into account possible uncertainties to the modeling of this function, we opted to allow $\{a_1,a_2\}$ to vary, introducing them among the Fisher parameters considered in the analysis (as described in section \ref{sec:MG}). Nonetheless, this inevitably introduces an extra source of uncertainty, disadvantaging predictions for the bright sources case and leading to forecasts in the Lensing-only case for bright sirens usually no better than those for dark sirens, as highlighted in the ``Lensing-only case'' subsection above. Nonetheless, one may wonder what the advantage of considering bright sources would be if the behaviour of $\Xi(z)$ was assumed fixed, getting rid of this extra source of uncertainty. Figure \ref{fig:ellipses_Lonly} provides constraints on $\mu_0, \eta_0, \Sigma_0$ (for $f_{\rm sky}=0.5$ and $T_{\rm obs}^{\rm GW}=15$yr) for the Lensing-only case, comparing the cases of $\{a_1,a_2\}$ open and $\{a_1,a_2\}$ fixed to fiducial values (with $w_0,\: w_a$ fixed). It shows a significant improvement in the constraining power of the experiments, with contour ellipses covering more reasonable ranges, highlighting a severe degradation in the constraining power due to the uncertainty on the modeling of the parameters describing $\Xi(z)$.

Indeed, Fisher estimated errors on $\{\mu_0, \eta_0, \Sigma_0\}$ when keeping $\{a_1,a_2\}$ fixed are the following: $\{1.32, \: 3.29,\: 0.70\}$ for the $\mathrm{GW \times IM}$ case and $\{0.75,\: 1.80,\: 0.08\}$ when adding galaxies. When comparing these numerical values to those in the ``LENSING'' rows of table \ref{tab:errors_darkbirght_NOw0wa}, we can see an improvement of up to one order of magnitude (for the $\mathrm{GW \times IM}$ case).

These results show that if the behaviour of $\Xi(z)$ was to be known, being able to detect an EM counterpart would be of crucial importance for experiments based only on the Weak Lensing observable, allowing to constrain $\{\mu_0, \eta_0, \Sigma_0\}$ with good accuracy, and significantly better than a case in which only dark sirens would be available. Nonetheless, an approach taking into account the uncertainty on the modeling of $\Xi(z)$ is safer and more realistic, although provides more pessimistic forecasts.

\subsubsection*{\textit{\{$\mathit{\mathbf{w_0,w_a}}$\} effects}}
Since we are studying theories with fixed background, it is natural to wonder about the impact of keeping the $\{w_0,w_a\}$ parameters fixed (results provided in table \ref{tab:errors_darkbirght_NOw0wa} and figures \ref{fig:musigma_ellipses_NOw0wacase}, \ref{fig:mueta_ellipses_NOw0wacase}) or open, as extra Fisher parameters (table \ref{tab:errors_darkbirght_withw0wa} and figures \ref{fig:musigma_ellipses_w0wacase}, \ref{fig:mueta_ellipses_w0wacase}). When fixing $w_0,w_a$ at their fiducial values results are in general either comparable or significantly more optimistic (up to a few factors unity), with smaller contour ellipses. As one would expect, the higher number of free parameters usually leads to less tight constraints.

\subsubsection*{\textit{$\mathit{\mathbf{\mathit{f_{sky}}}}$ effects}}
Improving the surveyed area of the sky logically improves the constraining power, sometimes significantly. It can be seen in figures \ref{fig:muSigma_ellipses_fsky_GWxIM_NOw0wacase}, \ref{fig:muSigma_ellipses_fsky_GWxIMxgal_NOw0wacase}, \ref{fig:mueta_ellipses_fsky_GWxIM_NOw0wacase}, \ref{fig:mueta_ellipses_fsky_GWxIMxgal_NOw0wacase} that the contours related to higher values of $f_{sky}$ (in magenta) are tighter than those for low $f_{sky}=0.1$ values (in green), sometimes reducing parameters degeneracies. This is valid for all considered probes: L, C and L+C (left, middle and right panels). We also report, not shown explicitly, a very mild dependence on the values of $T_{\rm obs}^{\rm GW}$, showing that in this framework the GW shot noise does not provide the bulk of the weight to the error budget.

\subsubsection*{\textit{Role of the EM counterpart for bright sirens}} As described over the course of this manuscript, the bright sirens case relies on the assumption that NSNS mergers are associated with an EM counterpart. This is an optimistic starting point, which is why we accompany these results to the BHBH+BHNS dark case. For completeness (although not explicitly reported here for the sake of brevity) we have also computed forecasts labelling all GW sources (BHBH, BHNS and NSNS mergers) as dark sirens. We found that results are generally comparable to the BHBH+BHNS dark case up to a few percentages. For this reason, results in this latter case can also be seen as a proxy for forecasts in a scenario characterized by a complete lack of EM counterparts.

\subsubsection*{\textit{Impact of $\ell_{max}$ for bright sirens}}
Throughout this section, we have provided results with a choice of $\ell_{\rm max}=300$ for detected bright sirens. As described in section \ref{sec:tracer_GW}, we are allowed to push our angular resolution limit beyond the intrinsic instrument limitation thanks to the detectability of EM counterparts. Nonetheless, we explored a set-up with an $\ell_{\rm max}=100$ even for bright sirens. This way, we are testing the extreme case in which EM counterparts would not be exploited for improving the angular resolution. Forecasts obtained this way are less optimistic than the $\ell_{\rm max}=300$ ones, with relative differences from just a few percentages (mainly for the Lensing-only case) up to a factor $\sim 5$ for the Clustering-only and L+C cases. Nonetheless, we note that this would not lead to orders of magnitude of difference among the forecasts, providing us fairly robust results to the specific $\ell_{\rm max}$ choice. 

\section{Conclusions}\label{sec:conclusions}
Cross-correlations between different tracers of the LSS and different observable probes can richly enhance the amount of physical information that can be extracted by present and forthcoming experiments. In this work we considered three different tracers: \textit{(i)} resolved GW signals from compact object mergers as observed by ET, both assuming the detection of EM counterparts (for NSNS, bright sirens) or not (for BHBH and BHNS, dark sirens); \textit{(ii)} the Intensity Mapping of the neutral hydrogen distribution as observed by the SKA-Mid survey; \textit{(iii)} resolved radio-galaxies as mapped by SKAO. This allows us to correlate and compare both GW and EM signals, testing the possible imprints of beyond-GR behaviours, as these two observables are supposed to respond in the same way to matter perturbations effects such as lensing. For this reason, the primary observational probe we took into account is the weak lensing power spectrum, both in auto and cross-tracers correlation. In order to gauge the effects of combining together different probes, we also introduced the angular clustering power spectra and their L$\times$C cross-term. We performed a Fisher matrix analysis in order to test a late-time parametrization scenario, forecasting the constraining power on the MG parameters $\{\mu_0,\eta_0,\Sigma_0\}$.

Our findings show that combining together different observational probes has a strikingly positive effect on the constraining power, with an improvement of up to an order of magnitude and results which are even competitive with constraints from Planck. We also find that, generally, cross-correlating together more tracers provides better constraints, as the combination of more information from different sources is more powerful than auto-correlation only experiments.  

In addition, we also show that when considering probes that describe physical effects that would be different between GW and EM sources (i.e., Lensing), the detection of an EM counterpart might be of crucial importance, allowing us to actively test the presence of different behaviours between these two observables and confirm or rule out GR alternatives to the description of gravity.

This work extends the efforts of the scientific community in the field of multi-tracing and multi-probes Astrophysics and Cosmology, showing that in an era rich in surveys and data (both from the present time and near-future experiments) the interconnection of different sources is able to yield results and constraints which are significantly more powerful than auto-correlation or single-probe results.

\section*{Acknowledgments}
We are thankful to Anna Balaudo, Nicola Bellomo, Lumen Boco, Giulia Capurri, Alice Garoffolo, Suvodip Mukherjee, Gabriele Parimbelli, Alvise Raccanelli, Marco Raveri, Marta Spinelli and the LSS group at IFPU for useful discussions. We thank the anonymous referee for thoughtful evaluation of our work. GS, MB and MV are supported by the INFN PD51 INDARK grant. MV is also supported by the ASI-INAF agreement n. 2017-14-H.0. AS acknowledges support from the NWO and the Dutch Ministry of Education, Culture and Science (OCW) (grant VI.Vidi.192.069).

%\clearpage

\appendix
\section{Constraints on $\{\mu_0, \eta_0\}$: plots}
\label{app:mueta}
In this appendix we provide contours plots on the constraints on the $\{\mu_0, \eta_0\}$ parameters. Comments on the results are embedded in the main text (section \ref{sec:forecasts}).

\begin{figure}[h!]
	\centering
	\hbox{\hspace{17mm}\includegraphics[width=0.75\linewidth]{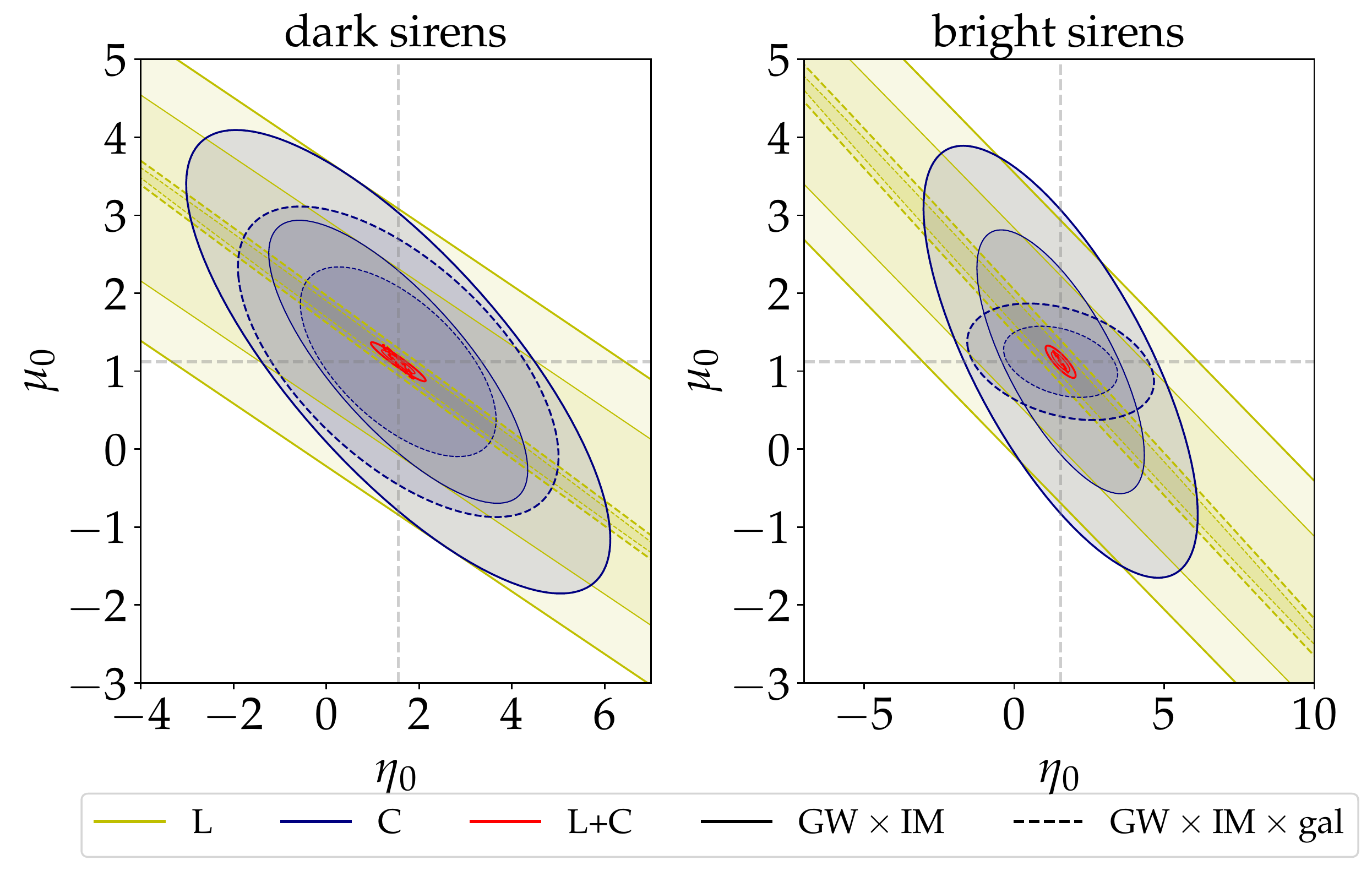}}
	\caption{Contours for $\{\mu_0,\eta_0\}$ in the GW$\times$IM (solid line) and GW$\times$IM$\times$gal (dashed line) cases, all probes considered (colors according to legend). Left panels refer to dark sirens (BHBH+BHNS), right panels refer to bright sirens (NSNS). $w_0, w_a$ are among the Fisher parameters considered. $T_{\rm obs}^{\rm GW}$=15 yr and $f_{\rm sky}=0.5$.}
	\label{fig:mueta_ellipses_w0wacase}
\end{figure}

\begin{figure}
	\centering
	\hbox{\hspace{16mm}\includegraphics[width=0.75\linewidth]{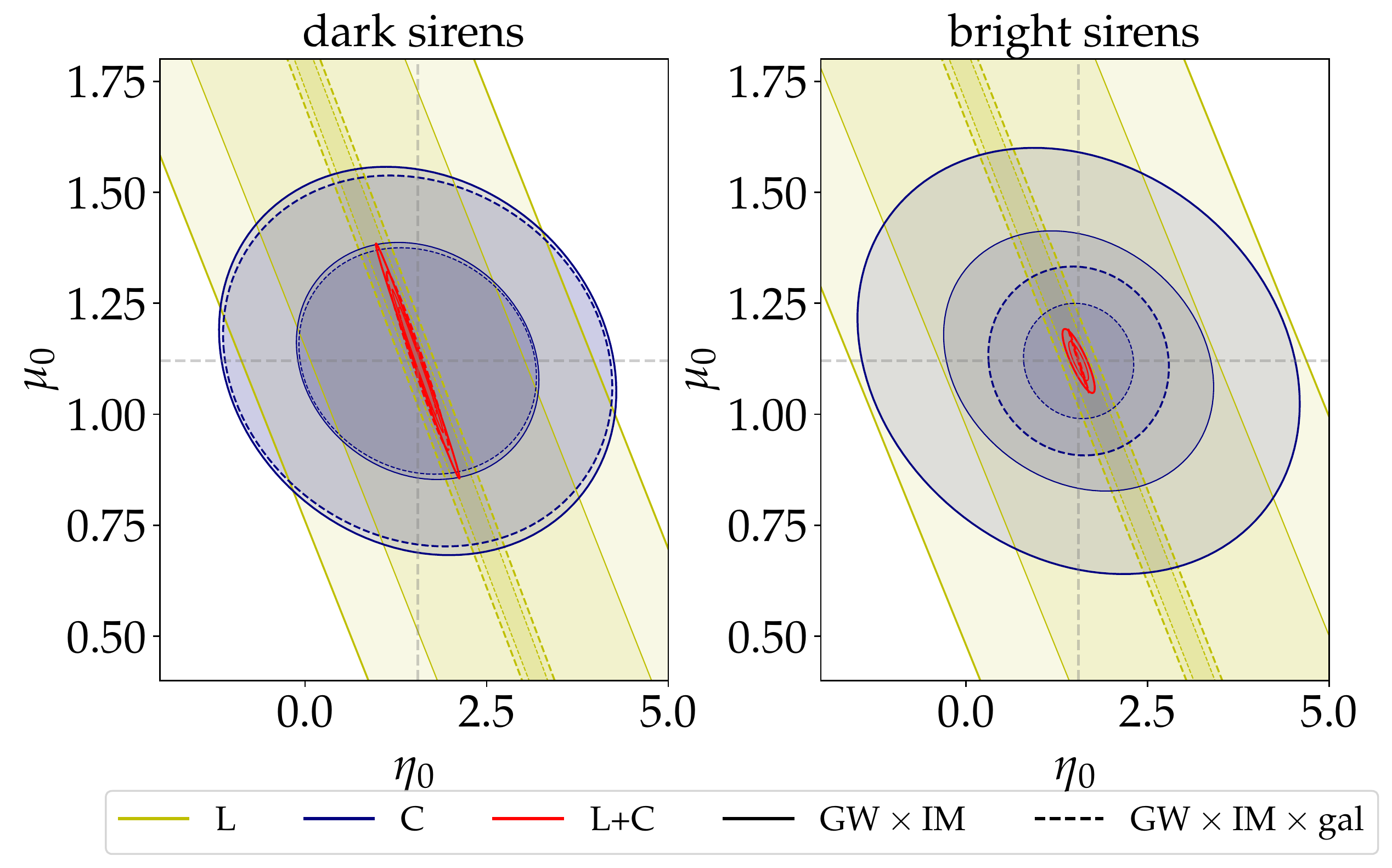}}
	\caption{Contours for $\{\mu_0,\eta_0\}$ in the GW$\times$IM (solid line) and GW$\times$IM$\times$gal (dashed line) cases, all probes considered (colors according to legend). Left panels refer to dark sirens (BHBH+BHNS), right panels refer to bright sirens (NSNS). $w_0, w_a$ are fixed to fiducial values. $T_{\rm obs}^{\rm GW}$=15 yr and $f_{\rm sky}=0.5$.}
	\label{fig:mueta_ellipses_NOw0wacase}
\end{figure}

\begin{figure}
	\centering
	\includegraphics[width=1.0\linewidth]{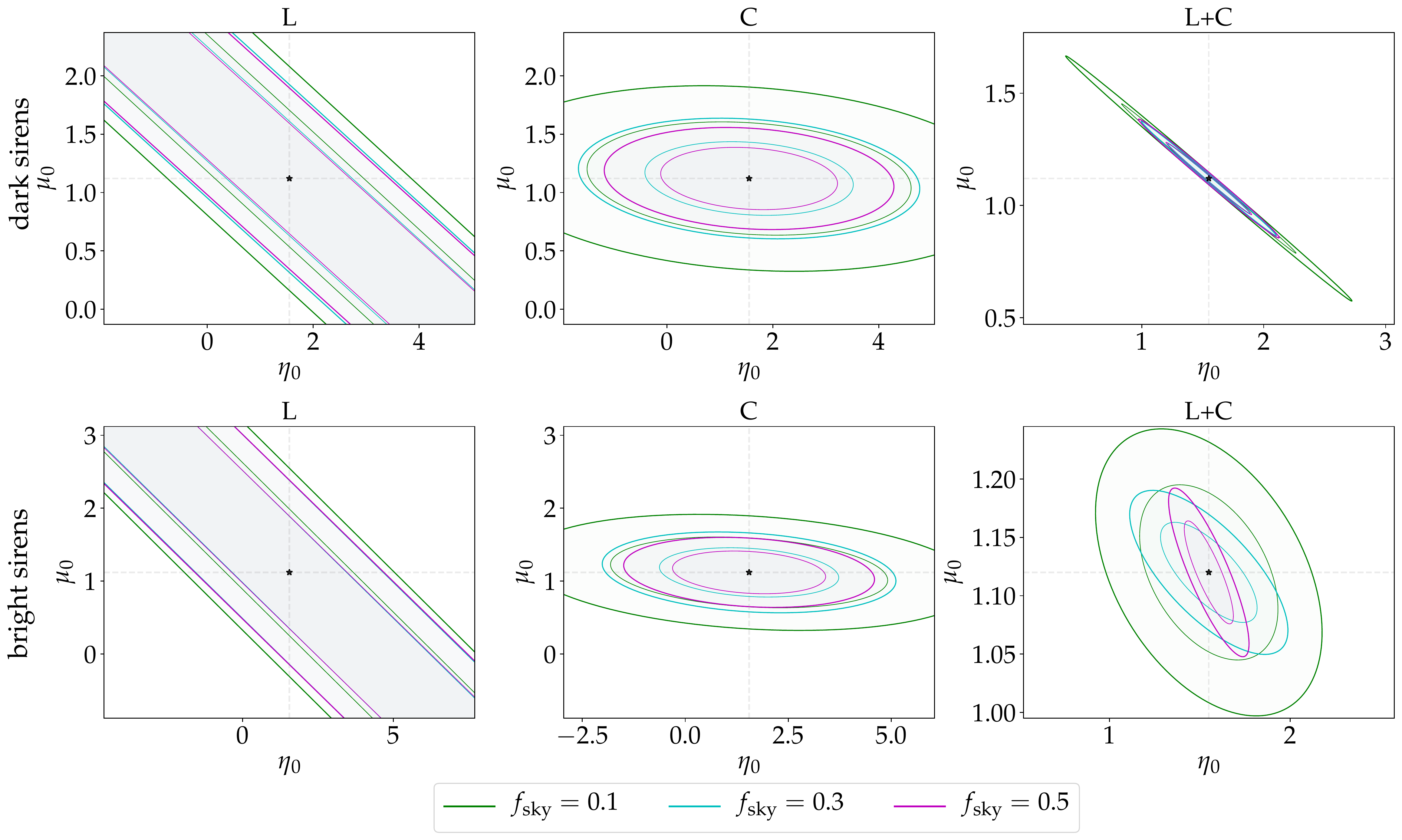}
	\caption{Contours for $\{\mu_0,\eta_0\}$ parameters for the GW$\times$IM case, all probes considered (only Lensing: left panels; only Clustering: center panels; Lensing + Clustering: right panels), for dark (top panels) and bright (bottom panels) sirens and for different values of $f_{\rm sky}$ (colour-coded according to legend). $w_0, w_a$ are fixed to fiducial values. $T_{\rm obs}^{\rm GW}$=15 yr.}
	\label{fig:mueta_ellipses_fsky_GWxIM_NOw0wacase}
\end{figure}

\begin{figure}
	\centering
	\includegraphics[width=1.0\linewidth]{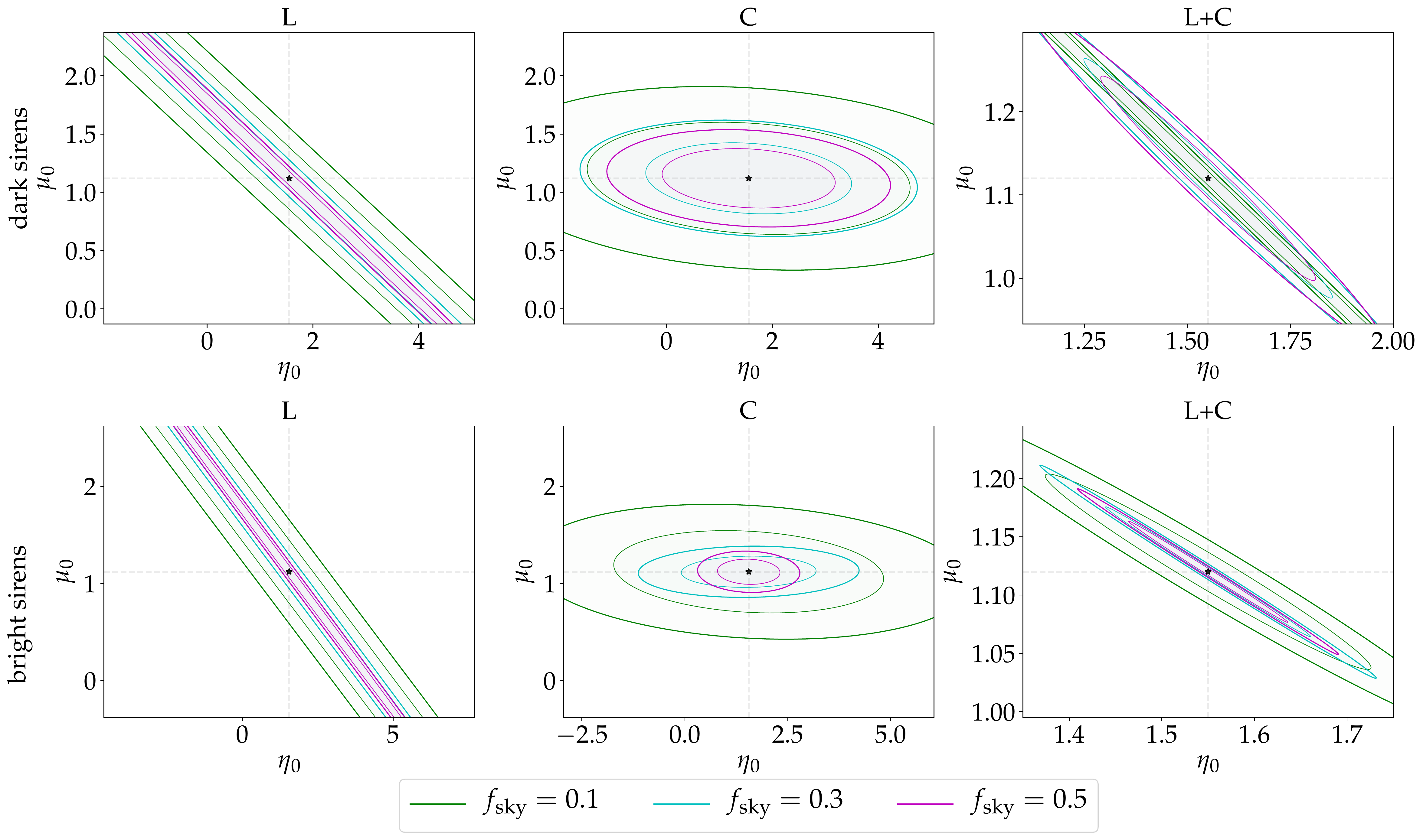}
	\caption{Contours for $\{\mu_0,\eta_0\}$ parameters for the GW$\times$IM$\times$gal case, all probes considered (only Lensing: left panels; only Clustering: center panels; Lensing + Clustering: right panels), for dark (top panels) and bright (bottom panels) sirens and for different values of $f_{\rm sky}$ (colour-coded according to legend). $w_0, w_a$ are fixed to fiducial values. $T_{\rm obs}^{\rm GW}$=15 yr.}
	\label{fig:mueta_ellipses_fsky_GWxIMxgal_NOw0wacase}
\end{figure}

\clearpage

\bibliography{biblio}
\bibliographystyle{utcaps}

\end{document}